\documentclass[a4paper,11pt]{article}
\pdfoutput=1 

\usepackage{jheppub} 

\usepackage{placeins}
\usepackage{subfig}
\usepackage{amsmath}
\usepackage{comment}
\usepackage{cleveref}
\usepackage{enumitem}

\usepackage{xspace}
\usepackage{main-defs}
\usepackage{rotating}
\usepackage{soul}
\usepackage{upgreek}
\usepackage{booktabs}
\usepackage{adjustbox}
\usepackage{pdflscape}
\usepackage{multirow}

\usepackage{longtable}

\usepackage{comment}

\usepackage{lineno}

\title{\boldmath Precision Measurements of Higgs Hadronic Decay Modes at the FCC-ee}

\author[a]{Andrea Del~Vecchio,}
\author[f]{Jan Eysermans,}
\author[d]{Loukas Gouskos,}
\author[c]{George Iakovidis,}
\author[e]{Alexis Maloizel,}
\author[e]{Giovanni Marchiori,}
\author[b]{and Michele Selvaggi}

\affiliation[a]{La Sapienza, Roma, Italy}
\affiliation[b]{European Organisation for Nuclear Research (CERN), Geneva, Switzerland}
\affiliation[c]{Brookhaven National Laboratory, Upton, NY, USA}
\affiliation[d]{Brown University, Providence, RI, USA}
\affiliation[e]{Laboratoire AstroParticule et Cosmologie, (CNRS/IN2P3), Paris, France}
\affiliation[f]{Massachusetts Institute of Technology, Boston, USA}

\emailAdd{andrea.del.vecchio@cern.ch}
\emailAdd{jan.eysermans@cern.ch}
\emailAdd{loukas.gouskos@cern.ch}
\emailAdd{george.iakovidis@cern.ch}
\emailAdd{alexis.maloizel@ens-paris-saclay.fr}
\emailAdd{giovanni.marchiori@cern.ch}
\emailAdd{michele.selvaggi@cern.ch}

\abstract{
The expected precision at the FCC-ee on the product $\sigma\times\mathcal{B}(H\rightarrow b\bar{b}, c\bar{c},s\bar{s},gg)$ of Higgs boson production cross sections times branching ratios of hadronic decays is presented. 
This study provides the first comprehensive determination of all major hadronic Higgs decay modes in a combined fit at future $e^+ e^-$ colliders, using both Higgs-strahlung ($ZH$) and Vector boson fusion ($\nu\bar{\nu} H$) production processes, with a full treatment of interference effects in the $\nu\bar{\nu} jj$ final state. 
It assumes four identical IDEA detectors collecting $e^+e^-$ collisions at $\sqrt{s}=240$ and $365\,$GeV. 
The combination of all channels across both energies, with full covariance between production and decay modes, yields a production cross-section times branching-ratio precision at the percent to per-mil level for the dominant hadronic final states ($b\bar{b}, c\bar{c},gg$). These results provide a comprehensive input to the determination of Higgs coupling projections at the FCC-ee, and they establish for the first time sensitivity to the rare decay $H\rightarrow s\bar{s}$, demonstrating that FCC-ee has the potential to provide evidence of the strange-quark Yukawa coupling.
}

\begin{document} 
\maketitle
\flushbottom

\section{Introduction}
\label{sec:intro}


The discovery of a Higgs boson (H) with mass $m_H\simeq125\,\mathrm{GeV}$ by the ATLAS and CMS experiments~\cite{HIGG-2012-27,CMS-HIG-12-028} was a landmark achievement in the understanding of the electroweak (EW) symmetry breaking~\cite{HIGG-2012-27,CMS-HIG-12-028}. 
All measurements so far~\cite{HIGG-2021-23,CMS-HIG-22-001,HIGG-2018-04,CMS-HIG-18-016,HIGG-2017-07,CMS-HIG-16-043,HIGG-2018-13,CMS-HIG-17-035,ATLAS:2025coj,CMS-HIG-19-006,HIGG-2022-22} (i.e., interactions with gauge bosons, third generation fermions, as well as rare decays to $Z\gamma$ and to a pair of muons)  show that its properties are consistent with those of the standard model (SM) Higgs boson~\cite{Englert:1964et,Higgs:1964ia}.

Precise measurements of the Higgs boson hadronic branching fractions are essential to test the mechanism of fermion mass generation in the SM. They provide direct access to the tree-level Yukawa couplings of the Higgs boson to bottom, charm, and strange quarks. In the SM, Yukawa couplings are proportional to the corresponding quark masses, and verifying this scaling is a fundamental test of the Higgs mechanism. A wide range of extensions of the SM predict possible deviations from this linear relation. Models such as Minimal Flavour Violation (MFV), strongly interacting or composite Higgs models, two Higgs doublet models (2HDM), supersymmetric extensions of the SM, and models with vector-like quarks can induce modifications of the tree level Yukawa couplings. A discussion of these models and their phenomenological implications can be found e.g. in Refs.~\cite{Cepeda:2019klc,Peskin:2022pfv}.

The first and second generations are especially informative: they directly test the mass–proportionality pattern far from the heavy-flavour regime and are uniquely sensitive to flavour-specific new dynamics. However, even at the High-Luminosity LHC (HL-LHC), most Higgs couplings are projected to reach at best few-percent precision, and direct probes of second-generation quark Yukawa couplings remain challenging; under optimistic assumptions, only an evidence-level sensitivity ($\sim\!3\sigma$) to $H\!\to\!c\bar c$ is anticipated~\cite{Cepeda:2019klc,CMS:2025hfp}. 
Moreover, these measurements rely on the assumption that the production cross-sections are equal to their SM-predicted values, thus introducing a degree of model dependence in the results.

In the SM, the hadronic branching ratios ($\mathcal{B}$) sum to about $69.32\%$. $\mathcal{B}(H\to s\bar{s})$ is expected to be around $2.4\times10^{-4}$, assuming $\bar m_s(2~\mathrm{GeV})=93~\mathrm{MeV}$~\cite{ParticleDataGroup:2024cfk}.
Therefore, precise measurements of the Higgs hadronic branching fractions are essential to test the mechanism of fermion mass generation in the SM. 
They give direct access to the tree-level Yukawa couplings to bottom, charm, and strange quarks, and constrain the effective, loop-induced coupling to gluons, which can be sensitive to new heavy coloured states.

The FCC-ee~\cite{FCC:2018evy,fsrPED}, a high-luminosity circular $e^+e^-$ collider, is expected to produce large Higgs samples in a clean environment at centre-of-mass energies around 240~GeV and above. 
At these energies, Higgs production is dominated by Higgs-strahlung ($e^+e^-\to ZH$), with cross-sections $\sigma_{ZH}\simeq200$~fb at $\sqrt{s}=240$~GeV and $\sigma_{ZH}\approx125$~fb at $\sqrt{s}=365$~GeV. The Vector-boson fusion (VBF) process also contributes at smaller rates: $0.4$ ($3.2$)~fb for $ZZ\to e^+e^-H$ and $6.1$ ($29.2$)~fb for $WW\to\nu_e\bar{\nu}_e H$ at $\sqrt{s}=240$ (365)~GeV. 
With $\mathcal{L}=10.8~\mathrm{ab}^{-1}$ at 240~GeV collected over three years with four detectors, about $2.2\times10^6$ Higgs-strahlung and $6.5\times10^4$ VBF events are expected. An additional five years at 365~GeV would yield about $370$k ($92$k) Higgs decays from $ZH$ (VBF) for $\sim3.12$~ab$^{-1}$.

At the FCC-ee, the absence of pile-up and underlying event, present in $pp$ collisions, and backgrounds dominated by EW processes, lead to much larger signal-to-background ratios than at the LHC, especially for hadronic decays. A further advantage is the recoil-mass technique, which measures the total Higgs production cross-section independently of the Higgs decay. Reconstructing each decay mode then enables direct branching-ratio measurements; a joint analysis of these results yields in a model-independent way, the total Higgs width and its couplings.

Previous studies of other $e^+e^-$ Higgs factories indicate sub-percent to few-percent precision on the leading hadronic modes. At the proposed Linear Collider Facilities (LCF), with $2.7~\mathrm{ab}^{-1}$ at $\sqrt{s}=250\,\mathrm{GeV}$, the projected statistical uncertainties on $\sigma_{ZH}\!\times\!\mathcal{B}(H\to b\bar b,\,c\bar c,\,gg)$ are about $0.41\%$, $2.5\%$, and $2.1\%$, respectively~\cite{LinearColliderVision:2025hlt}. At the proposed CEPC, projections with $20~\mathrm{ab}^{-1}$ at $\sqrt{s}=240$~GeV yield approximately $0.14\%$, $2.02\%$, and $0.81\%$ for the same quantities.

This article studies the sensitivity to $\sigma\times\mathcal{B}(\hbb,\,\cc,\,\ssbar,\,gg)$ at the FCC-ee, assuming four detectors with performance identical to those expected for the IDEA concept~\cite{IDEAStudyGroup:2025gbt}, and integrated luminosities of $10.8$ ($3.12$)~ab$^{-1}$ at $\sqrt{s}=240$ (365)~GeV. The primary production mode is $ZH$ with the $Z$ decaying to $e^+e^-$, $\mu^+\mu^-$, $\nu\bar{\nu}$, or quark–antiquark pairs of any flavour except $t\bar t$. The VBF production mode also contributes when the $Z$ is reconstructed as $e^+e^-$ (via $ZZ$ fusion) or as $\nu\bar\nu$ (via $WW$ fusion); the latter shares a final state with $Z(\nu\bar{\nu})H$ and is treated within the same analysis. The $Z$-fusion contribution is neglected due to its small rate except, in the $\ell\ell jj$ channel at $\sqrt{s}=365\,\mathrm{GeV}$ where it accounts for about $27\%$ of the total and is folded into the SM yields\footnote{At $\sqrt{s}=240\,\mathrm{GeV}$ the $Z$-fusion yield is about $3\%$.}. The $\ell\ell jj$ analysis is nevertheless optimised for $ZH$.

This work provides a coherent estimate of the precision in the determination of all major hadronic Higgs decay modes in a single combined fit at the FCC-ee, simultaneously accounting for both $ZH$ and VBF production and including a full treatment of interference effects in the $\nu\bar{\nu} jj$ final state. Measurements at $\sqrt{s}=240$ and $365$~GeV are combined, yielding the expected precision on $\sigma\times\mathcal{B}$ for $H\!\to\!b\bar b$, $c\bar c$, $s\bar s$, and $gg$ at both energies. Together with the FCC-ee determination of the total Higgs production cross-section~\cite{eysermans_2025_jfb44-s0d81}, these results provide essential input to model-independent global Higgs-coupling fits and to the extraction of the total Higgs width, expected to reach $0.78\%$ precision at the FCC-ee~\cite{fsrPED}. 
For the first time, the rare decay $H\!\to\!s\bar s$ is included consistently within the same combined fit, indicating a potential sensitivity to the strange-quark Yukawa coupling. The results 
have been provided to the European Strategy for Particle Physics Update (ESPPU2026)~\cite{selvaggi_2025_n78xk-qcv56} and to the Physics Briefing Book~\cite{deBlas:2025gyz}.

The article is organised as follows.
Section~\ref{sec:detector} summarises briefly the main features of the 
IDEA detector concept. The progress on the particle detection techniques that will allow building next-generation
devices that can perform efficient tagging of jets produced
by the hadronisation of gluons, charm quarks and
strange quarks is discussed. 
The simulated signal and background event samples
used for this study are described in Section~\ref{sec:samples}.
Section~\ref{sec:reconstruction} illustrates how
particle candidates and events are reconstructed.
The following three sections describe the event selection
and categorisation of the three final states under study, respectively $\ell\ell jj$,
$\nu\bar{\nu} jj$, and $jjjj$.
The measurement of $\sigma \times \mathcal{B}$ from the yields and
distribution of discriminant variables in the reconstructed event 
categories for each of the three analyses and their joint 
interpretation is discussed in Section~\ref{sec:globalfit}.

\section{The detector concept}
\label{sec:detector}

The results presented in this study are based on the projected performance of the IDEA detector concept~\cite{theideastudygroup2025ideadetectorconceptfccee}. The detector, roughly $11\times11\,\mathrm{m}$ in height and length respectively, consists of nested sub-detectors arranged around an FCC-ee interaction point.

The tracking system includes a silicon pixel vertex detector, surrounded by a low-mass drift chamber and an outer silicon wrapper. The vertex detector (VXD) includes a cylindrical barrel of five single layers at radii between $R=1.2$ and $31.5\,\mathrm{cm}$, and two endcaps (one per side of the IP), each made of three single-layer silicon disks. The VXD is designed to be ultra-light to minimise multiple scattering, and features an excellent single-point resolution of $\sigma = 3\,\mathrm{\upmu m}$. Its innermost layer is positioned at a radius of $13.7\,\mathrm{mm}$ few millimetres from the outer beam pipe radius of $11.7\,\mathrm{mm}$~\cite{Ilg_2025}. The resulting excellent impact-parameter resolution allows for efficient and pure identification of charm and bottom jets against those originating from gluons and light flavour jets.
The drift chamber (DC) has a total length of about $4\,\mathrm{m}$ and 112 wire layers at radii between 35 and $200\,\mathrm{cm}$, enabling continuous tracking and particle identification. The DC material traversed at $90^\circ$ is $\sim\!1.6\%\,X_0$. The silicon wrapper surrounds the DC, and provides a precise spatial point at the end of the trajectory and for measuring the charged-particle time of arrival. The tracker operates in a uniform $2\,\mathrm{T}$ axial magnetic field, provided by a thin superconducting solenoid covering radii between about 2.1 and $2.4\,\mathrm{m}$.

The combined particle-identification capabilities of the drift chamber and the silicon wrapper promise a $3\sigma$ $K/\pi$ separation over a broad momentum range. Time-of-flight from the silicon wrapper, assuming a timing resolution of about $100\,\mathrm{ps}$, ensures $K/\pi$ separation at low momenta, up to a few~GeV. At higher momenta, the separation is predominantly driven by ionisation energy-loss measurements in the drift chamber using the cluster-counting technique (d$N$/d$x$), which extends the $K/\pi$ discrimination up to momenta of about $30\,\mathrm{GeV}$. This high momentum particle-identification expected performance is essential for tagging kaons within heavy-flavour jets. In particular, the identification of leading kaons provides a distinctive signature of strange quark jets and plays a key role in the sensitivity to the $H\to s\bar{s}$ decay.

Outside the tracker, the electromagnetic and hadronic calorimeters, measure shower positions and energies and thus enables the reconstruction of photons, electrons, as well as charged and neutral hadrons. A thin solenoid is placed between them to preserve the excellent electromagnetic energy resolution. The $\sim 25\,\mathrm{cm}$-deep crystal electromagnetic calorimeter (ECAL) is based on lead tungstate crystals while organic scintillators are also considered. The hadronic calorimeter is based on fibre-sampling dual-readout calorimeter of about $2\,\mathrm{m}$ depth (corresponding to 7 interaction lengths) which measures electromagnetic and hadron showers. It features a stochastic term of $30\%/\sqrt{E}$ for hadrons, which would translate into an excellent visible energy and jet energy resolution, a critical aspect for the reconstruction of resonant Higgs hadronic decay modes. 

Finally, layers of $\upmu$-RWELL chambers embedded in the magnet return yoke surround the calorimeters and provide an efficient, cost-effective solution for identifying penetrating muons.

We note that the Higgs hadronic coupling measurements presented in this work rely on these detector performance assumptions. A degradation of the achievable $K/\pi$ separation would directly impact the strange-jet tagging efficiency and purity, and consequently the precision of the extracted strange-quark Yukawa coupling.

Beyond charged-hadron particle identification, the reconstruction of all hadronic Higgs decay modes depend critically on the assumed hadronic energy resolution and particle-flow performance of the detector. The impact of the assumed resolution is particularly pronounced for rare hadronic final states, such as $H\to c\bar{c}$ and $H\to s\bar{s}$, where the signal-to-background ratio is small. In addition, the excellent vertexing performance assumed for the detector plays a key role in the separation of $b$- and $c$-jets, and therefore directly affects the precision of the corresponding Higgs coupling measurements.

The impact of detector performance assumptions on hadronic Higgs coupling measurements, in particular those related to particle identification, vertexing, and hadronic energy resolution, has been extensively investigated in dedicated detector optimisation and performance studies for FCC-ee. An extensive discussion can be found in References~\cite{fsrPED, Abidi:2025mdw}.

\section{Event simulation}
\label{sec:samples}

Signal samples were generated with \wz~\cite{Kilian:2007gr}, interfaced to \py 6~\cite{Sjostrand:2006za} for Higgs boson decays and parton showering. 
Non-Higgs backgrounds were generated with \wz for processes producing purely leptonic final states, and for $W$-fusion $\nu_e\bar{\nu}_eZ$ production. \py 8~\cite{Sjostrand:2014zea} was used for processes producing hadrons in the final state.

Events were generated at both $e^+ e^-$ centre-of-mass energies of
$240\,\mathrm{GeV}$ and $365\,\mathrm{GeV}$. All samples are normalised to the expected integrated luminosities of $10.8$ ($3.12$)~ab$^{-1}$ at $\sqrt{s}=240~(365)\,\mathrm{GeV}$, using the highest available order theoretical predictions for their production cross-sections and branching ratios. 

A Gaussian smearing of the beam particle energy is applied to simulate the
momentum spread of the particles in the beam bunches.
The collision crossing angle is not taken into account in the simulation. 
The nominal IP corresponds to the nominal centre of the detector, which is also the origin of the detector coordinate system. A right-handed
coordinate system is used, with its origin at the nominal interaction point
in the centre of the detector and the \(z\)-axis along the beam pipe.
The \(x\)-axis points from the IP to the centre of the ring,
and the \(y\)-axis points upwards. Cylindrical coordinates \((r,\phi)\) are used in the transverse plane. 
The beamspot has an energy dependent longitudinal spread of $\sim1\,\textrm{mm}$ and transverse sizes of $\sim10\,\upmu$m ($x$) and $\sim30\,\textrm{nm}$ ($y$).

\wz was used for the production of Higgs boson event samples,
with the Higgs boson decaying to $b\bar{b}$, $c\bar{c}$, $gg$, $s\bar{s}$ or
the other dominant final states ($\tautau$, \ww, \zz).
\wz generates the kinematic distributions of the
Higgs boson and of the accompanying leptons, while the Higgs boson decay and
parton showering are handled by \py 6~\cite{Sjostrand:2006za}.
Both Higgs-strahlung and vector-boson-fusion production modes were included at $\sqrt{s}=240$ and $365\,\mathrm{GeV}$. Dedicated samples of Higgs boson decays to $b\bar b$, $c\bar c$, $s\bar s$, and $gg$ were also produced for training and validating the jet-tagging algorithms. These were generated independently of the analysis samples to avoid statistical correlations. \wz was also used for the production of non-Higgs background samples such as $ee\to ee$, $ee\to\mu\mu$, $ee\to\tau\tau$, and $ee\to \nu_e\bar{\nu}_eZ$ ($W$ boson fusion only, to avoid double-counting with $Z(\nu_e\bar{\nu}_e)Z$ events in the inclusive $ZZ$ sample), while \py 8 was used for the production of other non-Higgs backgrounds (inclusive \ww, \zz, \zqq production). In the latter, \py 8 is responsible for both the matrix element calculation and the decays of unstable particles.

To be able to study separately the Higgs-strahlung and VBF processes in the $\nu\bar{\nu} jj$ final state, samples of $ZH$, $Z\to \nu_\mu\bar{\nu}_\mu$ were also produced. These samples, scaled by three times their nominal value of $\sigma\times \mathcal{B}$ to account for the number of neutrino flavours, were used to calculate the expected distributions of kinematic variables for the pure Higgs-strahlung component, while the component from VBF and interference between the two processes, was estimated as difference between the distribution from the $ee\to \nu_e\bar{\nu}_e H$ simulation and that of $ee\to ZH, Z\to \nu_\mu\bar{\nu}_\mu$.

\section{Detector response and global event reconstruction}
\label{sec:reconstruction}

Generated events were passed through a parametric simulation of the detector response implemented in \del~\cite{deFavereau:2013fsa,Selvaggi:2014mya,Selvaggi:2016ydq}. Tracks are assumed to be reconstructed with 100\% efficiency for $p_T>100\,\textrm{MeV}$ within an angular acceptance of $100\,\textrm{mrad}$ in the IDEA tracking system. Track parameters are smeared according to the covariance matrix induced by the intrinsic detector resolution and by multiple scattering in the detector material~\cite{Bedeschi:2022rnj}. Electromagnetic and hadronic energy deposits are smeared in the dual-readout calorimeters according to their nominal expected resolution.

A particle-flow (PF) algorithm uses the charged track candidates
and the electromagnetic and hadronic energy deposits in the
calorimeter to define three lists of reconstructed
particle candidates: charged particles, photons and
neutral hadrons. Electrons and muons with $E>2\,\textrm{GeV}$ are assumed to be identified by dedicated algorithms with 99\% efficiency, and a similar efficiency is assumed for photons. Particle misidentification is neglected. The missing momentum vector ${\textrm{p}}^\mathrm{miss}$ is computed as the opposite of the vectorial sum of the momenta of all PF candidates. 

The partons produced by the Higgs boson hadronic decays
hadronise producing streams jets of hadrons
that interact with the detector.
A jet clustering algorithm is used to combine the reconstructed
particles into jet candidates. The jets' four-momenta are
then calculated as the sum of the four-momenta of the
jets' constituents.
Jets are reconstructed with the exclusive ``Durham'' $k_t$ algorithm~\cite{Catani:1991hj}, as implemented in \fj~\cite{Fastjet}. At each step, the two closest objects are merged until exactly $N$ jets remain, with $N=2$ for $\ell\ell jj$ and $\nu\bar\nu jj$ events, and $N=4$ for $jjjj$ events. The distance between two particles (or a particle and a jet)
is defined as:
\begin{equation}
d_{ij} = 2 \min (E_i^2, E_j^2)(1-\cos\theta_{ij}).
\label{eqn:dij}
\end{equation}
The $E$-recombination scheme is adopted, and jets
returned by the algorithm are sorted in decreasing
order of their energies.
The isolated leptons identified as the products of a 
$Z\to\ell\ell$ decay as described later are excluded from
the inputs for the clustering.

To distinguish the flavour of the parton initiating
a hadronic shower reconstructed as a jet, a tagging
algorithm is used. This is based on a graph neural network algorithm exploiting various properties of the particle-flow inputs and their relationships clustered into the jets. The main features of the algorithm are detailed in Ref.~\cite{Bedeschi:2022rnj}, though a more recent version~\cite{aumiller_2024_8g834-jv464} using a Transformer architecture is employed in this study

The network outputs seven scores, normalised to unity, corresponding to the probabilities that a jet originates from a $b$, $c$, $s$, $u$, or $d$ quark, from a gluon, or from a hadronic decay of a $\tau$ lepton. 
The network was trained on independent simulated $Z(\nu\bar\nu)H$ events, clustered into two jets with the same $N=2$ Durham algorithm as in the analysis. The true flavour label of each jet was derived from the generated Higgs decay.
\FloatBarrier

\section{\texorpdfstring{The $\ellell jj$}{lljj} final state}
\label{sec:llh}

The analysis of the $\ell\ell jj$ ($\ell=e,\mu$) final state focuses on Higgs-strahlung events. While the analysis is optimised in $ZH$ events, the contribution from $Z$-boson fusion is present, and represents $<3\%$ of the total number of expected Higgs events at $\sqrt{s}=240\,\mathrm{GeV}$ and $\sim27\%$ at $\sqrt{s}=365\,\mathrm{GeV}$ before the event selection. This channel has the best signal-to-background ratio among the three under study but the smallest expected yields, due to $\mathcal{B}(Z\to\ell\ell) = 6.7\%$.

\subsection{Event selection}
\label{ssec:llh_selection}

Events are required to contain two leptons (electrons or
muons)\footnote{charge conjugation is implicitly assumed}
of the same flavour and opposite-sign  charges.
The lepton momenta are required to be between the kinematic limits imposed by two-body kinematics and the $Z$ boson boost in 
$e^+e^-\to ZH$ events: $25-80\,\mathrm{GeV}$ ($13-160\,\mathrm{GeV}$) at $\sqrt{s}=240~(365)\,\mathrm{GeV}$. 

In case of multiple same-flavour, opposite-charge lepton pairs satisfying the previous
requirements, the one with invariant mass closest to the nominal
$Z$ boson mass of $91.2\,\mathrm{GeV}$ is selected. 
The selected di-lepton pair is required to have 
invariant mass within a $20\,\mathrm{GeV}$-wide range ($81-101\,\mathrm{GeV}$)
around the nominal $Z$ boson mass.

In $ZH$ signal events, the $Z$ boson is expected to recoil against
a Higgs boson. By exploiting energy-momentum conservation and the knowledge of the initial 4-momentum of the collision, the recoil mass $m_\mathrm{recoil}$
can be calculated as:
\begin{equation}
  m_\mathrm{recoil}^2 = (\sqrt{s} - E_{l\bar{l}})^2 - p_{l\bar{l}}^2 
\end{equation}

where $P_{\ell\ell} =
(E_{\ell\ell}, \vec{p}_{\ell\ell})$
is the reconstructed four-momentum of the di-lepton pair.
In $ZH$ events, the recoil mass is expected to be peaked near the Higgs boson mass.
Events are therefore required to pass the requirement
$120<m_\mathrm{recoil}<140$~GeV.

After the lepton candidates from the $Z\to\ell\ell$ decays have
been identified, jets are reconstructed by clustering all the
p-flow candidates after excluding the two selected leptons.

A veto for events containing additional leptons with momenta above $25\,\mathrm{GeV}$
and a requirement that the variables $d_{2,3}$ and $d_{3,4}$
are
greater than zero is applied to suppress background
events from Higgs decays to non-hadronic final states,
such as $H\to\tau\tau$ with one
of the $\tau$ leptons decaying to leptons, from $ZZ$ events
and from inclusive $Z(\ell\ell)$ production.

The selection criteria, and the expected yields and efficiencies of each
selection step for signal and background processes, are listed
in Table~\ref{tab:llh_cutflow} for $\sqrt{s}=240\,\mathrm{GeV}$ (Table~\ref{tab:llh_365_cutflow} shows the yields at $\sqrt{s}=365\,\mathrm{GeV}$ in Appendix~\ref{appendix}).

The final selection efficiencies for $\ell\ell H$ events at $\sqrt{s}=240\,\mathrm{GeV}$
are 60\%, 62\%, 63\%, 63\% for Higgs bosons decaying to
$b\bar{b}$, $c\bar{c}$, $g\bar{g}$, or $s\bar{s}$, respectively.
The efficiency for other Higgs boson decays is 46\% for $WW$,
54\% for $ZZ$ and 41\% for $\tau\tau$;
when only fully hadronic decays of the Higgs boson
decay products are considered, the efficiency is 63\% for $WW$, 59\% for $ZZ$ and 56\% for $\tau\tau$.

$Z$-boson fusion events are estimated to contribute, after the selection, 
to only about 3\% of the total signal.

At $\sqrt{s}=365\,\mathrm{GeV}$, the final efficiencies are around
50\%, 54\%, 54\%, 54\% for $\ell\ell H$ events  with the Higgs boson
decay respectively to $b\bar{b}$, $c\bar{c}$, $g\bar{g}$, or $s\bar{s}$, and 32\%, 38\%, 20\% for the $WW$, $ZZ$ and \tautau decays. Considering only
fully hadronic decays for these last three cases, the efficiencies are
55\%, 51\% and 47\%, respectively.

\begin{sidewaystable}[!htbp]
  \centering
  \caption{
    Selection criteria, expected yields and efficiency for signal and background processes for the \texorpdfstring{$\ellell jj$}{lljj} analysis, 
    assuming an integrated luminosity $\mathcal{L}=10.8$~\iab of $e^+e^-$ collisions
    at $\sqs = 240$~GeV. The equivalent yield at $\sqs = 365$~GeV are shown in Appendix~\ref{appendix}, Table~\ref{tab:llh_365_cutflow}.
  }
  \label{tab:llh_cutflow}
  \resizebox{\textwidth}{!}{
  \begin{tabular}{lrrrrrrrrrrrrrrrrrrrrrrrr}
\toprule
Selection                   & \multicolumn{2}{c}{$\ell\ell H(b\bar{b})$} & \multicolumn{2}{c}{$\ell\ell H(c\bar{c})$} & \multicolumn{2}{c}{$\ell\ell H(gg)$} & \multicolumn{2}{c}{$\ell\ell H(s\bar{s})$} & \multicolumn{2}{c}{$\ell\ell H(WW)$} & \multicolumn{2}{c}{$\ell\ell H(ZZ)$} & \multicolumn{2}{c}{$\ell\ell H(\tau\tau)$} & \multicolumn{2}{c}{$ZZ$} & \multicolumn{2}{c}{$WW$} & \multicolumn{2}{c}{$Z/\gamma^*(\ell\ell)$} & \multicolumn{2}{c}{$Z/\gamma^*(q\bar{q})$}  \\    
                             & Yield & \effpct &  Yield & \effpct &  Yield  & \effpct &  Yield & \effpct &  Yield & \effpct &  Yield & \effpct & Yield & \effpct &  Yield   & \effpct &  Yield     & \effpct &  Yield    & \effpct &  Yield     & \effpct \\
\midrule
No cuts                      & 87599 &      -  &  4348  &      -  &   12313 &      -  &    36  &     -   &  32368 &      -  &   3971 &      -  &  9433 &      -  & 14677092 &       - &  177535800 &      -  & 146804400 &       - &  568662120 &       - \\
one $Z(\ell\ell)$ candidate  & 78728 &     90  &  3906  &     90  &   11058 &     90  &    32  &    90   &  29314 &     91  &   3594 &     91  &  8546 &     91  &  1223530 &       8 &    1667859 &      1  &  28920864 &      20 &     304657 &       0 \\
\mll in 81--101~GeV          & 70832 &     90  &  3514  &     90  &    9948 &     90  &    29  &    90   &  26247 &     90  &   3244 &     90  &  7662 &     90  &   784799 &      64 &     410769 &     25  &  12264100 &      42 &      16264 &       5 \\
\mrec in 120--140 GeV        & 67630 &     95  &  3356  &     96  &    9502 &     96  &    28  &    96   &  24948 &     95  &   3059 &     94  &  7281 &     95  &    96931 &      12 &     179156 &     44  &   2014189 &      16 &      3161  &      19 \\
no leptons with $p>25$~GeV   & 64531 &     95  &  3325  &     99  &    9495 &    100  &    28  &   100   &  18326 &     73  &   2757 &     90  &  5398 &     74  &    87009 &      90 &     178913 &    100  &   2014187 &     100 &      2839  &      90 \\
$\djj[2,3]>0$, $\djj[3,4]>0$ & 64531 &    100  &  3325  &    100  &    9495 &    100  &    28  &   100   &  18136 &     99  &   2612 &     95  &  4781 &     89  &    69857 &      80 &       8072 &      5  &     71760 &       4 &      2839  &     100 \\
\midrule
$\ell=e$                     & 31730 &      -  &  1633  &      -  &    4664 &      -  &    14  &     -   &   8904 &      -  &   1284 &      -  &  2369 &      -  &    35366 &       - &       5048 &      -  &     67285 &       - &      1425  &       - \\
$\ell=\mu$                   & 32801 &      -  &  1692  &      -  &    4831 &      -  &    14  &     -   &   9232 &      -  &   1328 &      -  &  2412 &      -  &    34491 &       - &       3024 &      -  &      4475 &       - &      1414  &       - \\
\bottomrule
  \end{tabular}
  } 
 \end{sidewaystable}

\begin{figure}[!htbp]%
  \centering
  \subfloat[]{\includegraphics[width=0.49\textwidth]{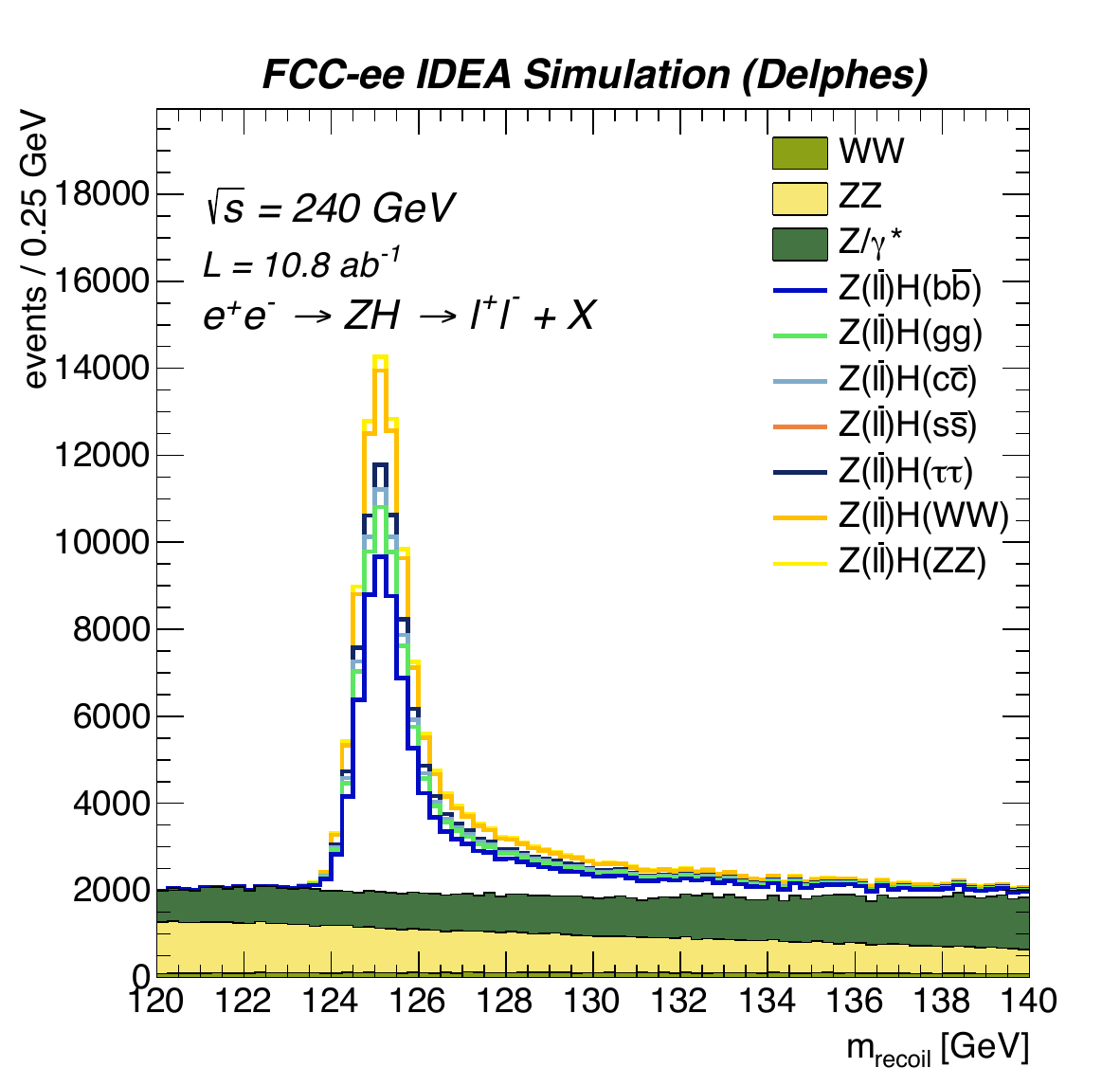}}
  \subfloat[]{\includegraphics[width=0.49\textwidth]{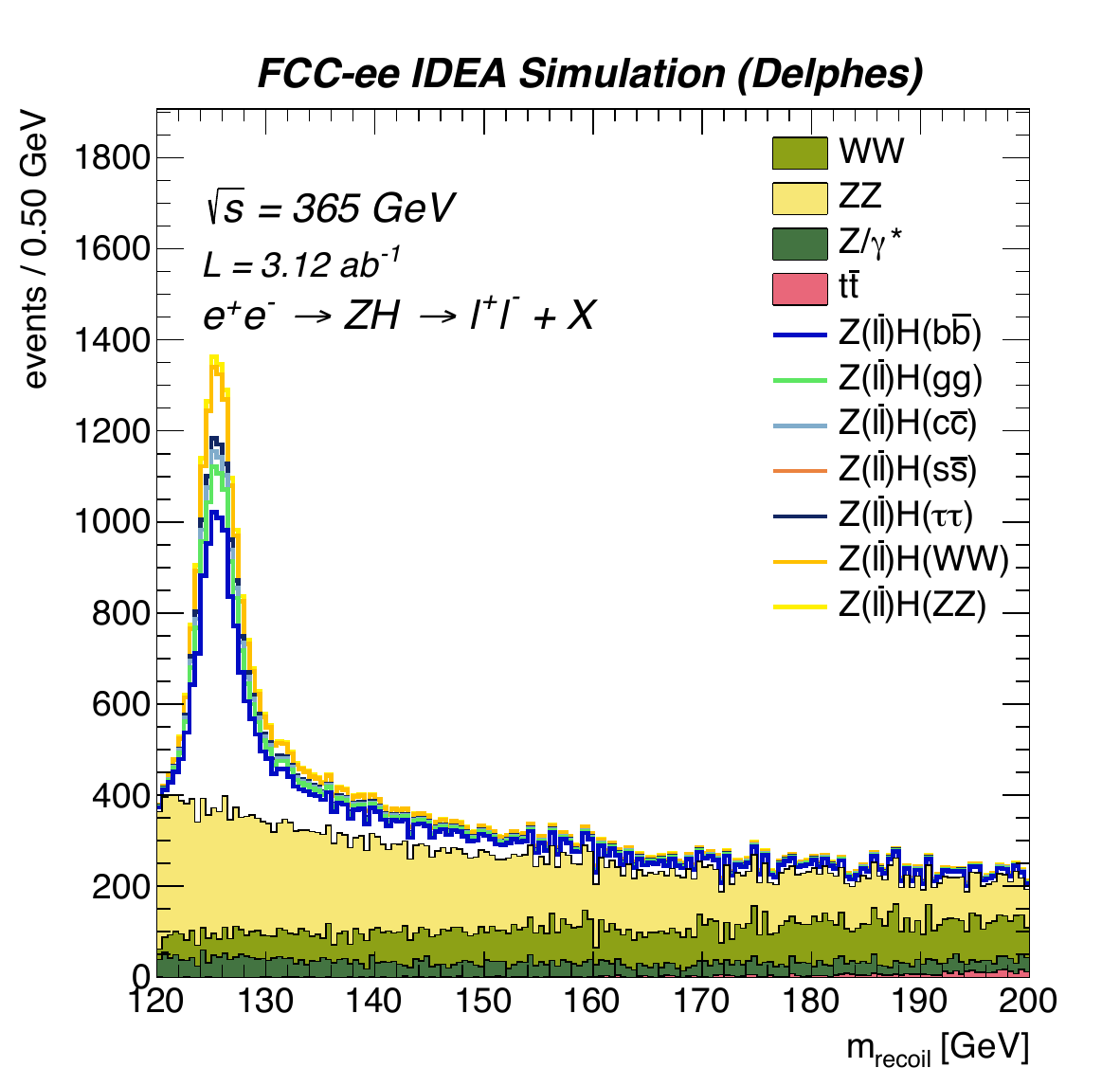}}
  \caption{Recoil mass distribution of signal and background superimposed
    after the final selection for (a) $\sqrt{s}=240\,\mathrm{GeV}$ and (b) $\sqrt{s}=365\,\mathrm{GeV}$.}
\label{fig:llh_mrecoil}
\end{figure}

The expected recoil mass distributions for signal and
background processes superimposed after
the selection are shown in Figure~\ref{fig:llh_mrecoil}. At $\sqrt{s}=365\,\mathrm{GeV}$, the higher centre-of-mass energy, larger beam energy spread and larger lepton momenta lead to a 
wider signal distribution with a longer tail extending to higher values of the recoil mass. For a similar reason, the tail of the $Z$ boson peak from $Z(\ell\ell)Z(q\bar{q})$ events extends to significantly higher values of \mrec at $\sqrt{s}=365\,\mathrm{GeV}$, leading to a worse signal-to-background ratio S/B than at $\sqrt{s}=240\,\mathrm{GeV}$.

\subsection{Event classification}
\label{ssec:llh_categories}

After the selection described in the previous section, events
at $\sqrt{s}=240~(365)\,\mathrm{GeV}$
are classified in ten (eleven) mutually orthogonal categories corresponding to the
seven Higgs boson decays -- to \bb, \cc, \ssbar, $gg$, $WW$, $ZZ$ and $\tau\tau$ -- and the three (four) main background processes: $ZZ$, $WW$, $Z/\gamma^*(\ell\ell)$ (and $t\bar{t}$ at $\sqs=365$~GeV).
The classification is performed by a fully connected neural network
(NN) with a number of input nodes equal to the number
of input variables, four hidden layers (with respectively 128, 128, 128, and 64 nodes), and ten (eleven) output nodes corresponding to the 
target processes.
The training of the neural network is performed with Keras~\cite{Chollet2015Keras}
using the Tensorflow~\cite{tensorflow2015-whitepaper} backend.
For the training, a subset of events is selected for each
Higgs boson decay category from the signal samples, and for the background categories from the background samples.
Two thirds of the events are used to train the network, and the other one third is 
used to validate it.
Once trained, the network is evaluated on the full samples of
Section~\ref{sec:samples}.

The input variables used by the neural networks are the seven
output scores $K_j = (b_j,c_j,s_j,g_j, \tau_j, u_j, d_j)$, $j = 1, 2$ of the flavour tagging algorithm~\cite{Bedeschi:2022rnj,aumiller_2024_8g834-jv464} for each of the two jets in the selected events (14 inputs), the di-jet invariant mass,
the missing momentum, and the jet clustering distance variables $d_{2,3}$ and $d_{3,4}$. 

\begin{figure}[!htbp]%
  \centering
    \subfloat[]{\includegraphics[width=0.49\textwidth]{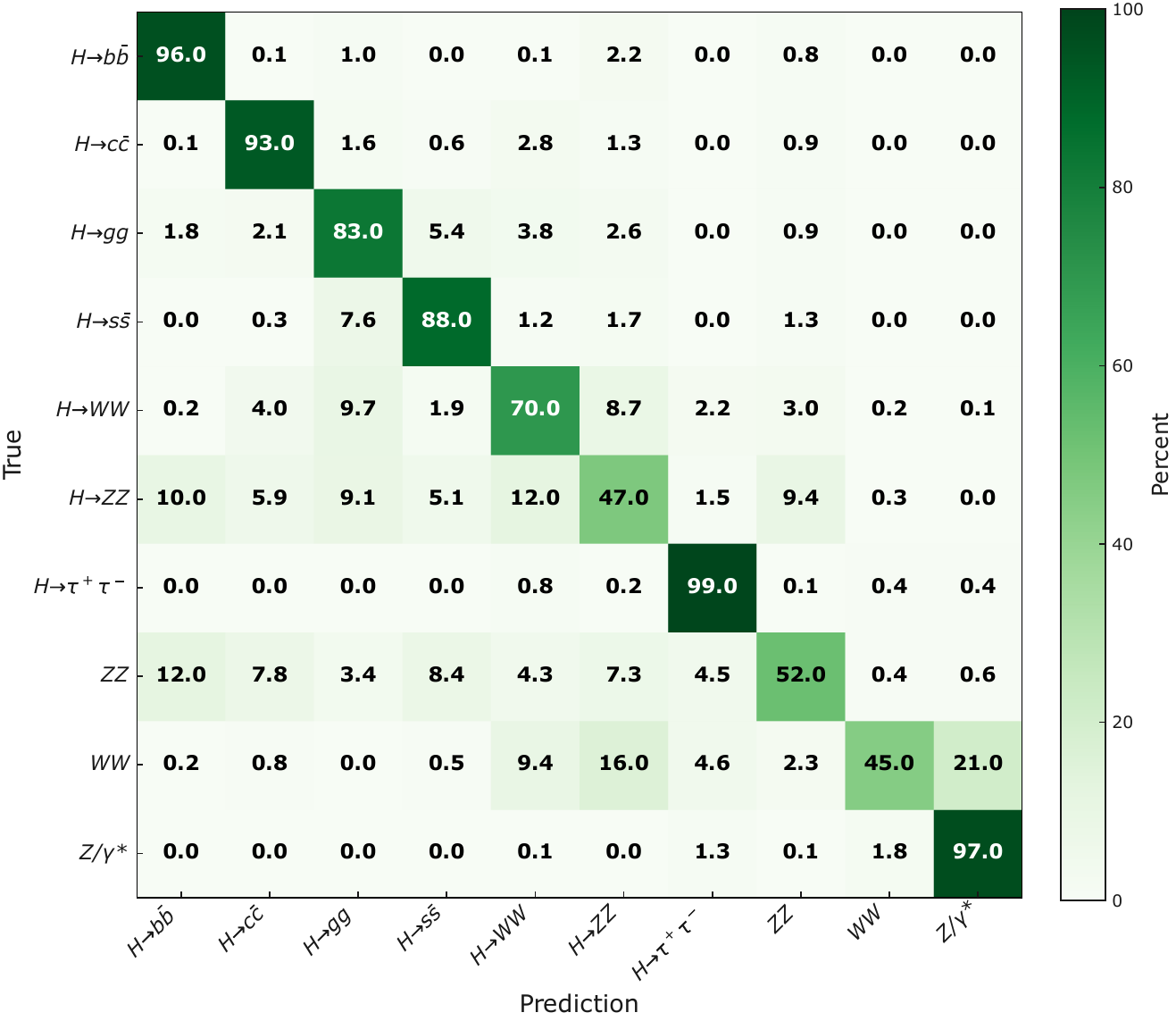}}
    \subfloat[]{\includegraphics[width=0.49\textwidth]{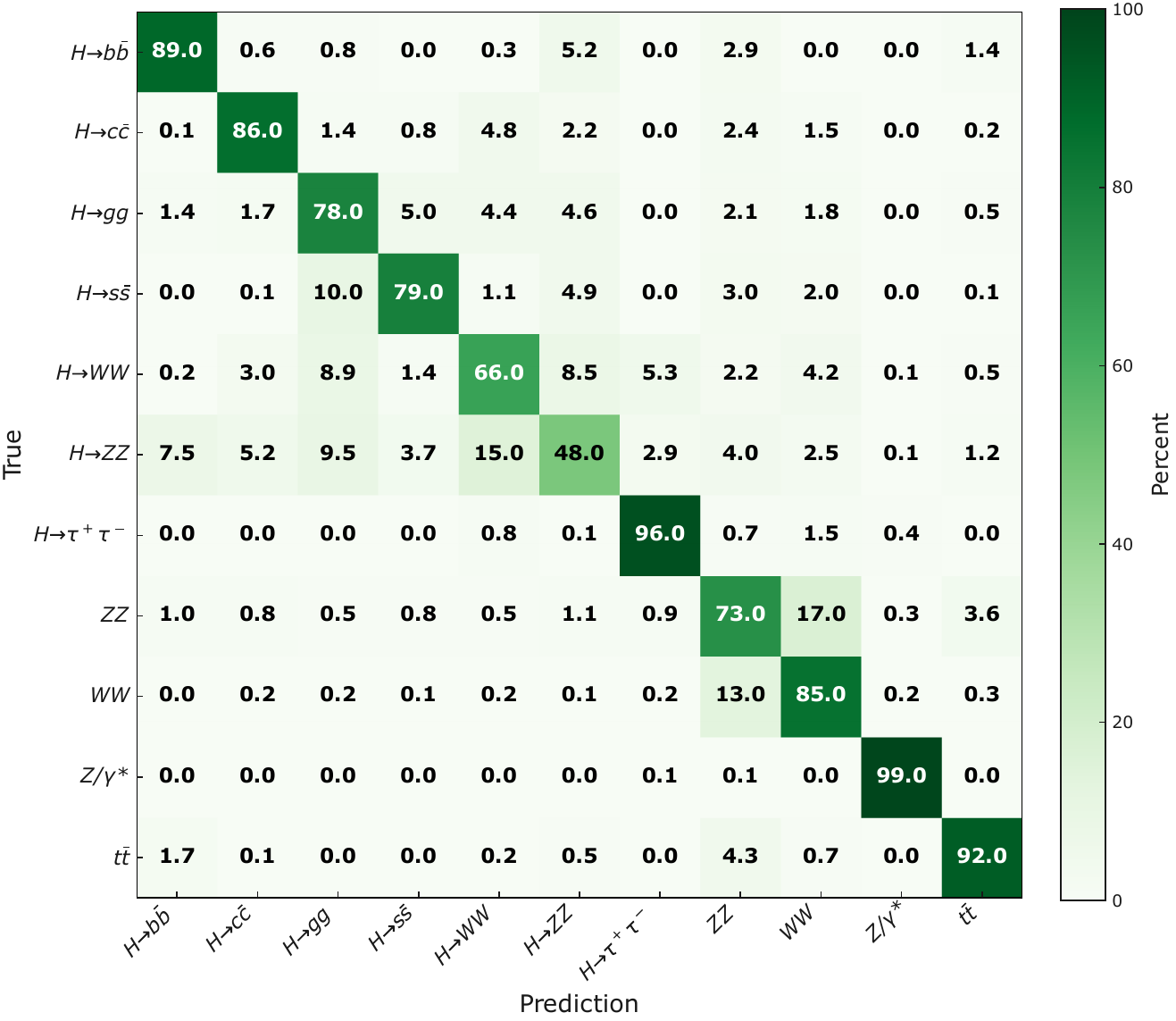}}
  \caption{Confusion matrix of the NN after the final training, evaluated on
    the validation subsample, for (a) $\sqs=240$~GeV and (b) $\sqs=365$~GeV.}
\label{fig:llh_nn_training}
\end{figure}

In addition, the cosine of the polar angle $\theta_{\ell\ell}$ of the di-lepton pair is also included as an input of the neural network, to exploit the difference in angular distributions of the leptons from
$Z\to \ell\ell$ decays in $ZZ$ and $ZH$ events due to the 
different spins of the $Z$ ($J=1$) and Higgs ($J=0$) boson~\cite{Barger:1993wt}.
The training is performed over 50 epochs, after which the performance tends to saturate.
The events are labelled as $ZH(XX)$ $XX=b\bar{b}, c\bar{c}, s\bar{s}, gg, WW, ZZ, \tau\tau$ or 
$ZZ$, $WW$, $Z/\gamma^{*}$ (and $t\bar{t}$ at $\sqrt{s}=365$~GeV) based on the output node of the NN with the largest score.
The overall accuracy of the trained network is 80\%, and no overtraining is observed.
The migration matrices evaluated on the validation sample of the training are shown in
Figure~\ref{fig:llh_nn_training}.
The confusion among the target processes is slightly larger
at $\sqrt{s}=365\,\mathrm{GeV}$ due to the addition of one more target,
the higher beam energy spread leading to worse signal/background separation.

To increase the final sensitivity of the analysis, the seven $ZH(XX)$ 
categories are further split into up to three subcategories based on the value of the NN output score of the events. The thresholds used to define the categories are chosen in order to maximise the expected combined significance summing in quadrature $S/\sqrt{S+B}$, where $S$ and $B$ are the signal and background yields in each category, leading to a
total of up to 24 (25) categories at $\sqs=240~(365)\,\mathrm{GeV}$.

\section{The $\nu\bar{\nu} jj$ final state}
\label{sec:nunuh_ii}
The analysis of the $\nunu jj$ final state aims at extracting both $ZH$ (with $Z\to\nunu$) and VBF Higgs boson production events.
This channel has the second best signal-to-background ratio among the three under study and an expected yield before selection that is about three times larger than the $\ell\ell jj$ final state, due to $\mathcal{B}(Z\to\nunu) = 20\%$.
The reconstructed final state contains two jets from the Higgs boson decays  and large momentum imbalance due to the neutrinos  from the $Z$ boson decay. Hence, for signal events, the reconstructed di-jet invariant mass $m_{jj}$ is expected to peak at the Higgs boson mass value, while the recoil mass of the di-jet system $m_\mathrm{recoil}$ is expected to peak at the $Z$ boson mass value in Higgs-strahlung production and to accumulate at higher values in VBF events.

\subsection{Event selection}
\label{ssec:nunuh_selection_ii}

Events containing at least one lepton with $p>20$ GeV are vetoed. This requirement removes leptonic or semi-leptonic background events and ensures orthogonality with respect to the selection of the $\ell\ell jj$ analysis. 

A loose preselection on the di-jet invariant mass $m_{jj}>20\,\mathrm{GeV}$ is applied, in addition to a requirement on the recoil mass $m_\mathrm{recoil}>10\,\mathrm{GeV}$ that ensures orthogonality with the analysis of fully hadronic final states described in the next section. The cosine of the angle between the visible momentum and the beam axis is required to satisfy the requirement $\abscosthjj<0.85$ to suppress single and di-boson backgrounds. sFurthermore, requirements on $d_{1,2}$ and $d_{2,3}$ suppress events with a topology consistent with more than two well separated jets. The number of expected events after each step of the selection for signal and background processes is shown in Table~\ref{tab:pre_selection_ii}.

\begin{table}[!htbp]
	\centering
    \footnotesize
	\caption{Yields after each step of the selection and final efficiency for each signal and background process for the $\nu\bar{\nu}jj$ analysis for VBF and $ZH$ production modes, assuming an integrated luminosity $\mathcal{L}=10.8$~\iab of $e^+e^-$ collisions at $\sqs = 240$~GeV. The equivalent yield table for $\sqs = 365$~GeV is shown in the Appendix~\ref{appendix} in Table~\ref{tab:pre_selection_365_transformed}.}
    \begin{tabular}{l|cccc|c} \hline
          &   Yield  & Yield  & Yield after & Yield after \\
        Process &  before   & after & $\vert \mathrm{cos} \theta \vert<0.85$ & kinematics $\&$ d$_{ij}$ & Efficiency($\%$)\\
         & selection & lepton veto & requirement & requirements & \\
        \hline
    $\nu_e\bar{\nu}_eH, H(b\bar{b})$ & $3.80\times10^{4}$ & $3.80\times10^{4}$ & $3.49\times10^{4}$ & $2.86\times10^{4}$ & 75.3 \\
    $\nu_e\bar{\nu}_eH, H(c\bar{c})$ & $1.90\times10^{3}$ & $1.90\times10^{3}$ & $1.86\times10^{3}$ & $1.53\times10^{3}$ & 80.5 \\
    $\nu_e\bar{\nu}_eH, H(gg)$       & $5.40\times10^{3}$ & $5.40\times10^{3}$ & $5.40\times10^{3}$ & $4.40\times10^{3}$ & 81.5 \\
    $\nu_e\bar{\nu}_eH, H(s\bar{s})$ & $1.57\times10^{1}$ & $1.57\times10^{1}$ & $1.58\times10^{1}$ & $1.28\times10^{1}$ & 81.5 \\
    $\nu_e\bar{\nu}_eH, H(\tau\tau)$ & $3.60\times10^{3}$ & $3.60\times10^{3}$ & $2.59\times10^{3}$ & $2.00\times10^{3}$ & 55.6 \\
    $\nu_e\bar{\nu}_eH, H(WW)$       & $1.37\times10^{4}$ & $1.37\times10^{4}$ & $9.70\times10^{3}$ & $7.80\times10^{3}$ & 56.9 \\
    $\nu_e\bar{\nu}_eH, H(ZZ)$       & $1.64\times10^{3}$ & $1.64\times10^{3}$ & $1.44\times10^{3}$ & $1.14\times10^{3}$ & 69.5 \\

    $\ZH, H(b\bar{b})$ & $2.52\times10^{5}$ & $2.52\times10^{5}$ & $2.31\times10^{5}$ & $5.40\times10^{4}$ & 79.2 \\
    $\ZH, H(c\bar{c})$ & $1.25\times10^{4}$ & $1.25\times10^{4}$ & $1.22\times10^{4}$ & $1.05\times10^{4}$ & 84.3 \\
    $\ZH, H(gg)$       & $3.54\times10^{4}$ & $3.54\times10^{4}$ & $3.54\times10^{4}$ & $3.06\times10^{4}$ & 86.4 \\
    $\ZH, H(s\bar{s})$ & $1.04\times10^{2}$ & $1.04\times10^{2}$ & $1.04\times10^{2}$ & $8.97\times10^{1}$ & 86.2 \\
    $\ZH, H(\tau\tau)$ & $2.37\times10^{4}$ & $2.37\times10^{4}$ & $1.71\times10^{4}$ & $1.37\times10^{4}$ & 57.7 \\
    $\ZH, H(WW)$       & $9.00\times10^{4}$ & $9.00\times10^{4}$ & $6.36\times10^{4}$ & $5.40\times10^{4}$ & 60.0 \\
    $\ZH, H(ZZ)$       & $1.08\times10^{4}$ & $1.08\times10^{4}$ & $9.54\times10^{3}$ & $7.80\times10^{3}$ & 72.2 \\
    $qqH$                       & $5.76\times10^{5}$ & $5.76\times10^{5}$ & $4.92\times10^{5}$ & $2.60\times10^{4}$ & 3.84 \\
    $WW$                        & $1.73\times10^{8}$ & $1.73\times10^{8}$ & $1.06\times10^{8}$ & $1.37\times10^{7}$ & 7.51 \\
    $ZZ$                        & $1.40\times10^{7}$ & $1.40\times10^{7}$ & $1.10\times10^{7}$ & $2.67\times10^{6}$ & 18.8 \\
    $Z/\gamma^* \to q\bar{q}$  & $5.66\times10^{8}$ & $5.66\times10^{8}$ & $5.47\times10^{8}$ & $5.85\times10^{6}$ & 1.01 \\
        \hline
    \end{tabular}

  \label{tab:pre_selection_ii}
\end{table}

\begin{figure}[!htbp]
\centering
\subfloat[]{\includegraphics[width=0.49\textwidth]{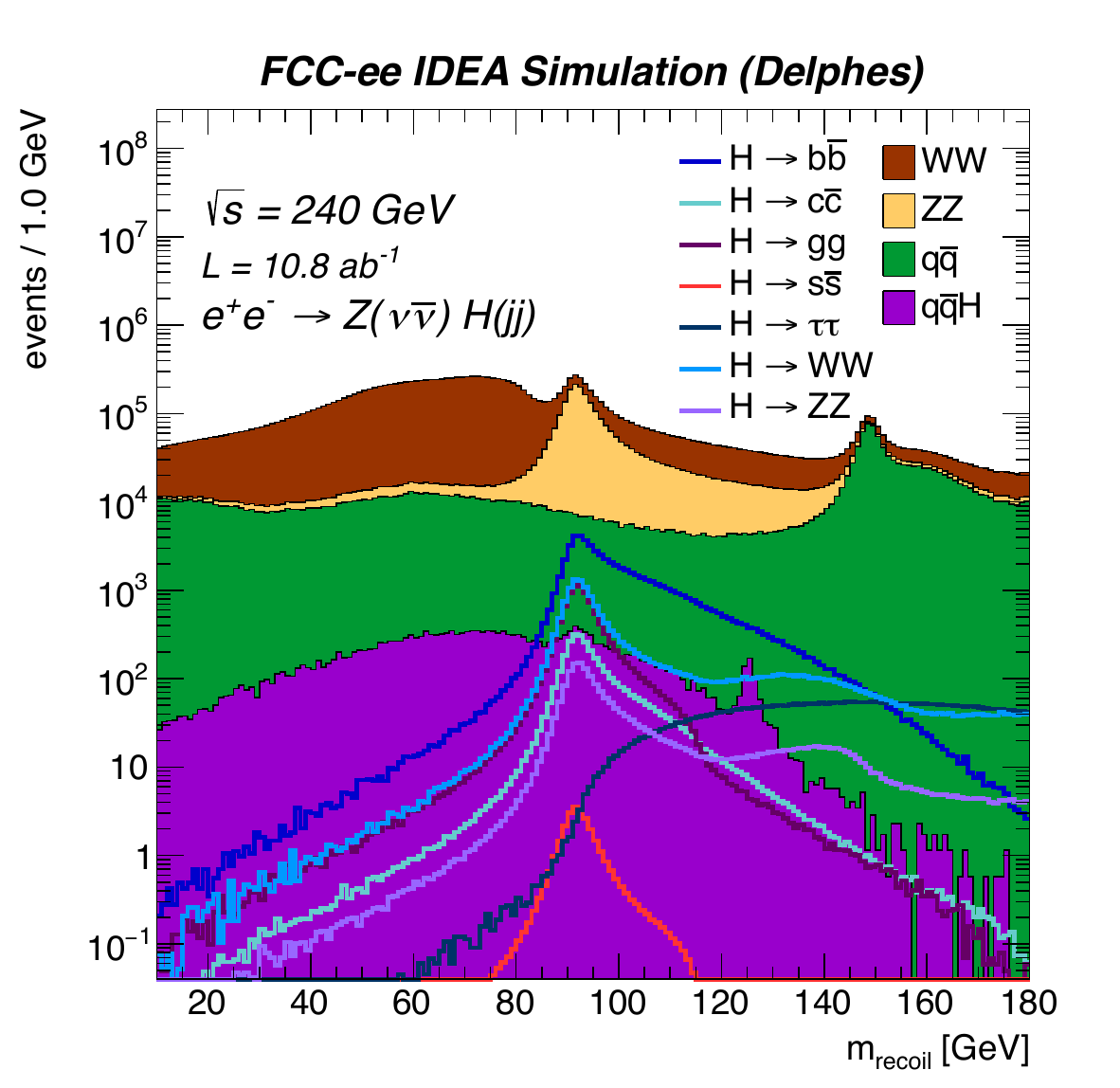}}
\subfloat[]{\includegraphics[width=0.49\textwidth]{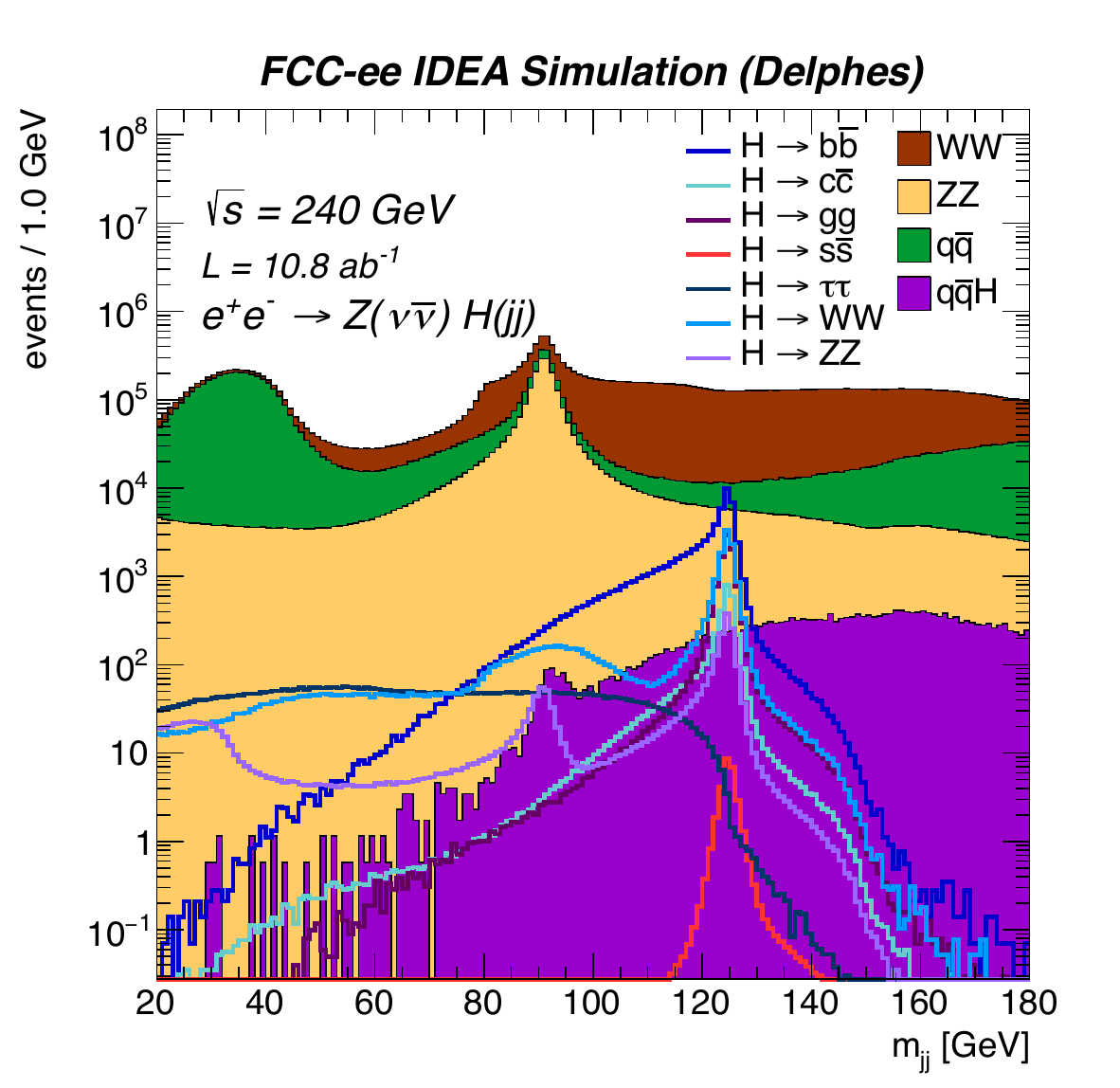}}  \\
\subfloat[]{\includegraphics[width=0.49\textwidth]{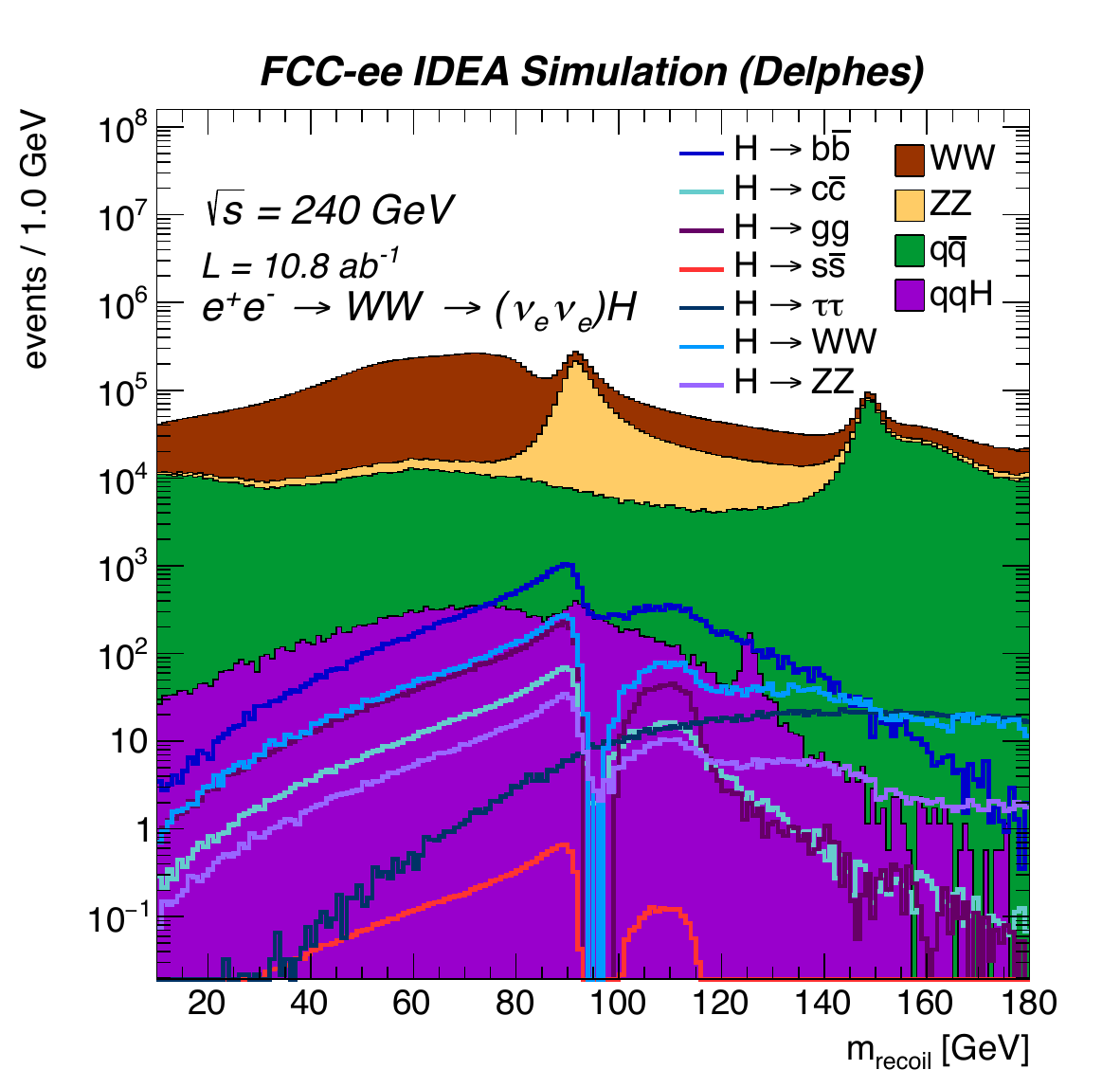}}
\subfloat[]{\includegraphics[width=0.49\textwidth]{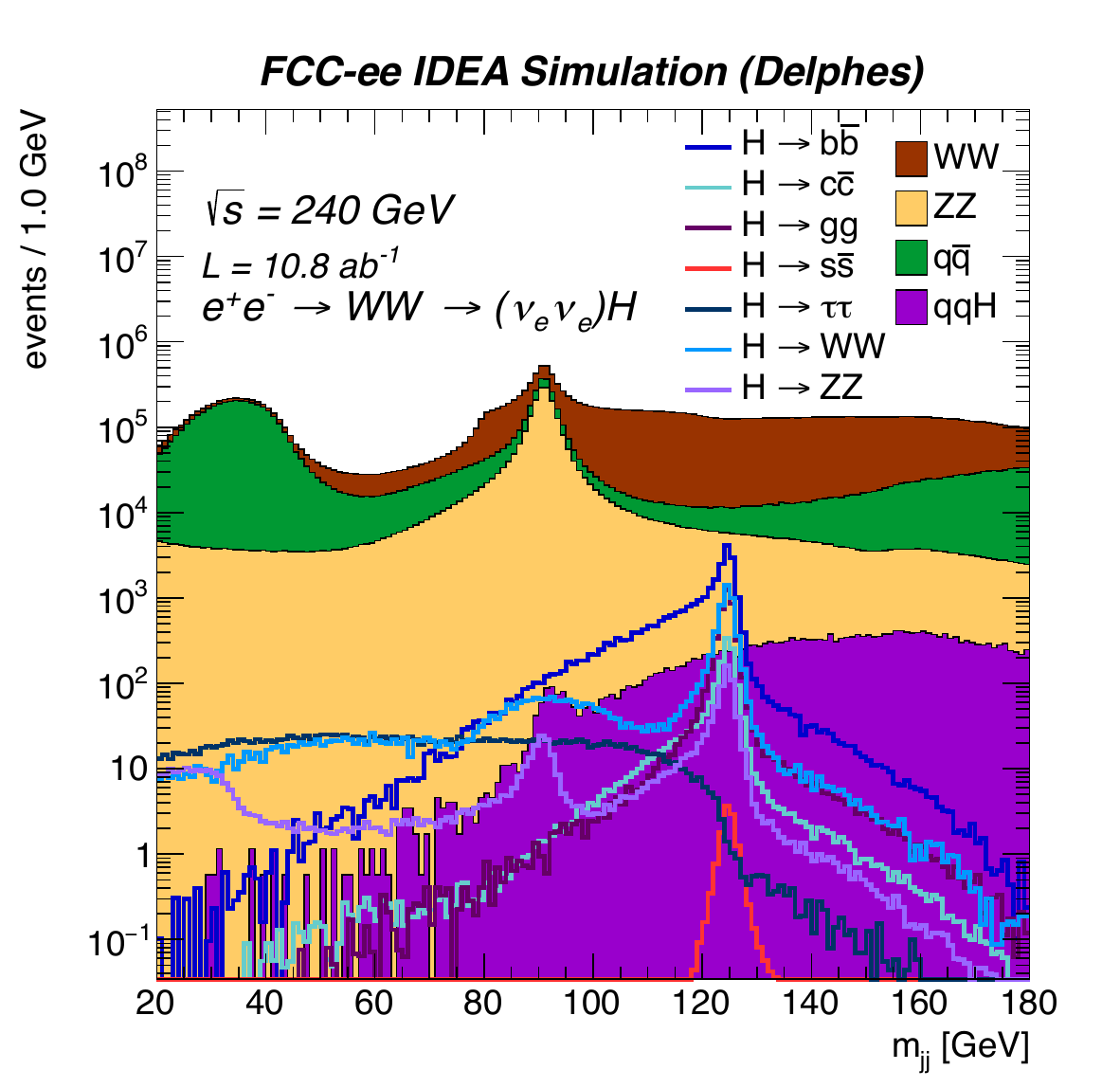}}
\caption{The $m_{\mathrm{recoil}}$ and the di-jet invariant mass $m_{jj}$ distributions after pre-selection for the $ZH$ production mode (a,b) and for the VBF production mode (c,d) at $\sqrt{s}=240\,\textrm{GeV}$.}
\label{fig:Mjjpre-aftersel}
\end{figure}

\begin{figure}[!htbp]
\centering
\subfloat[]{\includegraphics[width=0.49\textwidth]{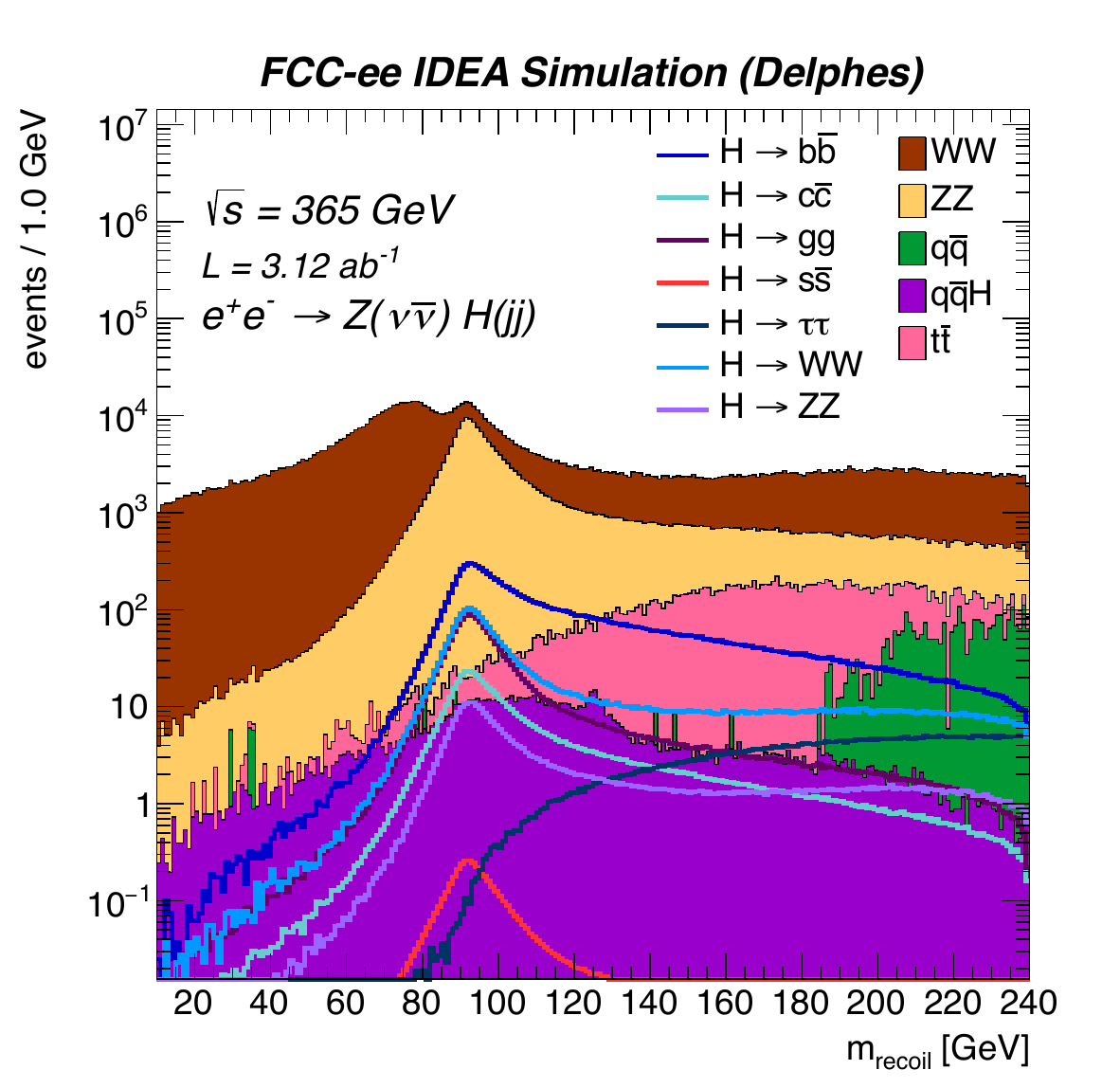}}
\subfloat[]{\includegraphics[width=0.49\textwidth]{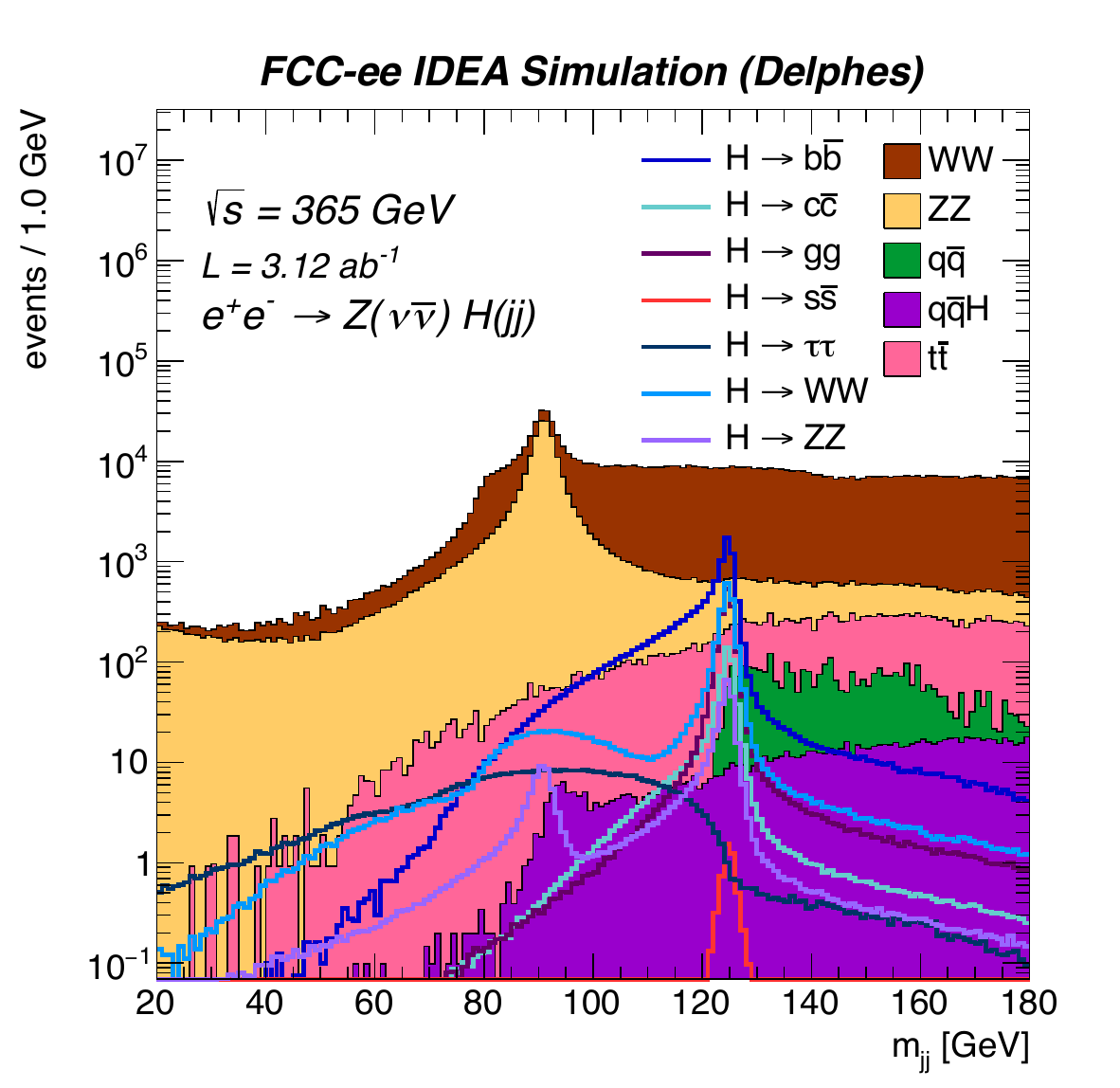}}\\
\subfloat[]{\includegraphics[width=0.49\textwidth]{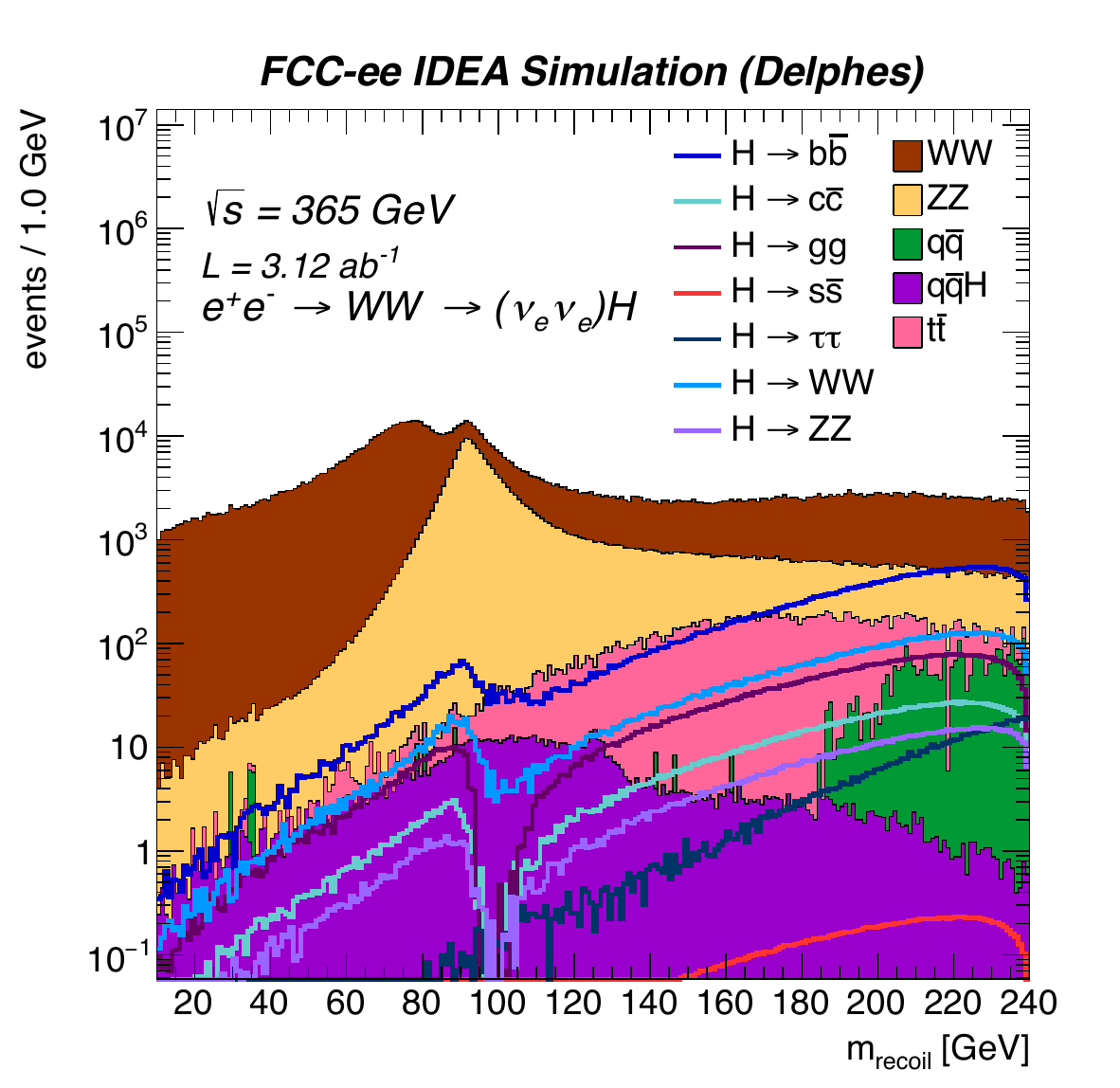}}
\subfloat[]{\includegraphics[width=0.49\textwidth]{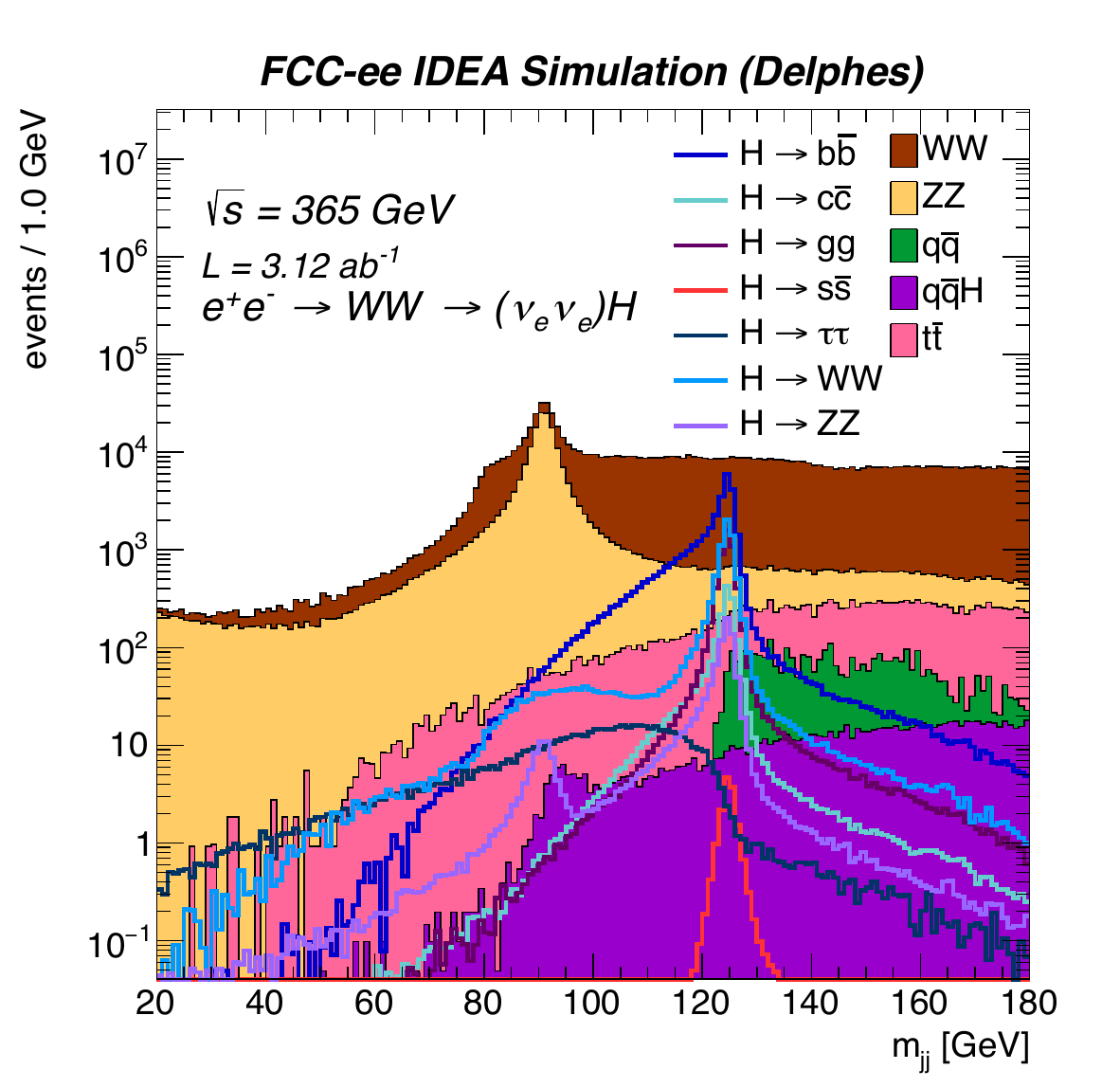}}
\caption{The $m_{\mathrm{recoil}}$ and the di-jet invariant mass $m_{jj}$ distributions after pre-selection for the $ZH$ production mode (a,b) and for the VBF production mode (c,d) at $\sqrt{s}=365\,\textrm{GeV}$.}
\label{fig:Mjjpre-aftersel-365}
\end{figure}

Figure~\ref{fig:Mjjpre-aftersel} (\ref{fig:Mjjpre-aftersel-365}) 
shows the recoil mass and the di-jet invariant mass distributions of signal and background events passing the pre-selection at $\sqrt{s}=240\,\mathrm{GeV}$ ($\sqrt{s}=365\,\mathrm{GeV}$).

\subsection{Event classification}
\begin{figure}[ht]
  \begin{center}
  \includegraphics[width=0.7\textwidth]{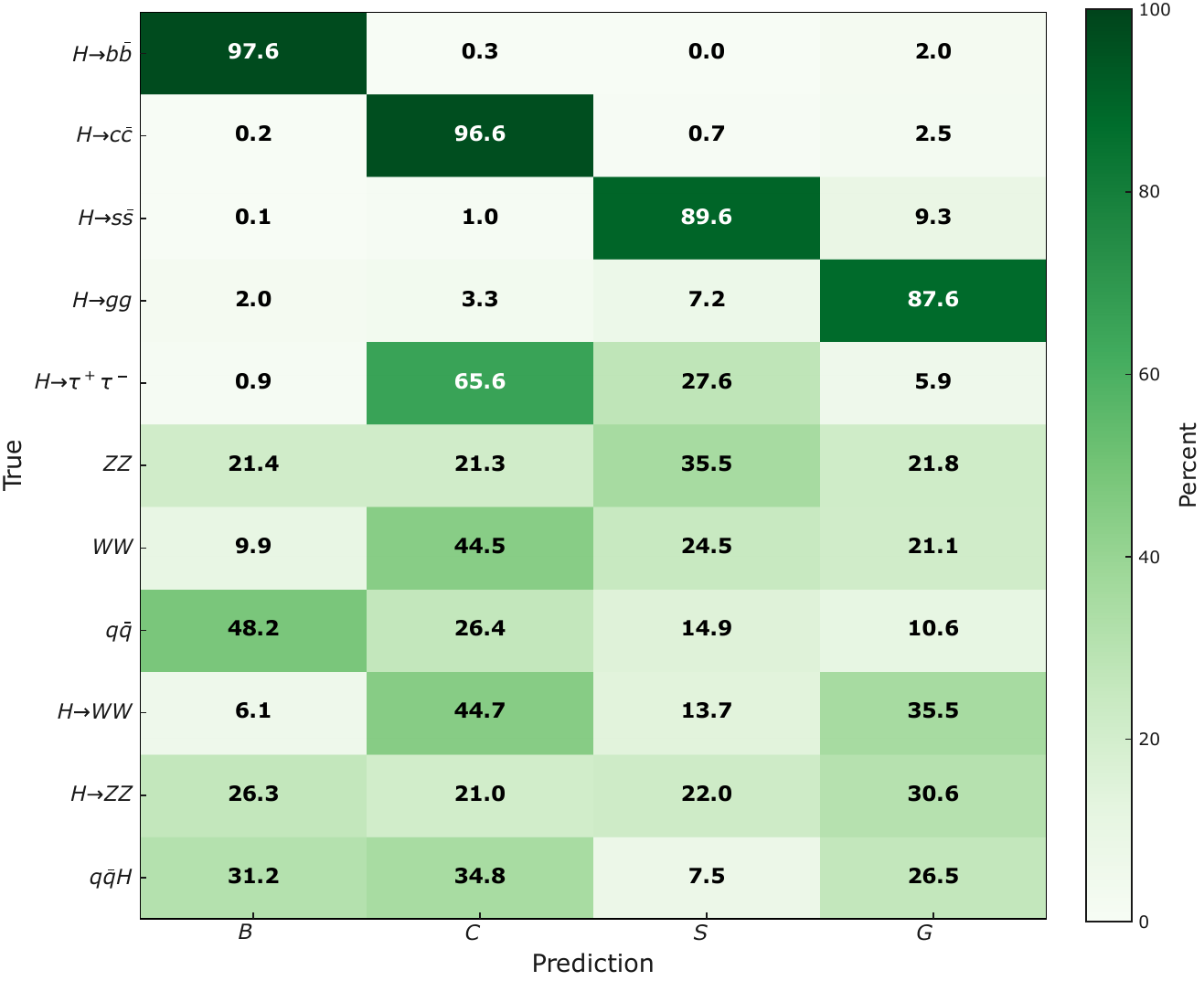}
  \caption{Migration matrix in the four categories of interest, for the $\nunu jj$ analysis at $\sqrt{s}=240$~GeV.
  The numbers correspond to the fraction of events (in percent) of a given process (vertical axis) after the preselection, that are classified in one of the four $K$-like categories (horizontal axis).}
  \label{fig:confusion_matrix_ii}
  \end{center}
  \end{figure}

Seven event-wise flavour-tagging observables $K = K_1 + K_2$, $K=b,c,s,g,\tau, u, d$, are calculated from the sum of the flavour probabilities $K_j$ ($j=1,2)$ of the two reconstructed jets as determined by the jet flavour tagging algorithm. 
The $K$-scores are then used to classify the events into seven orthogonal "$K$-like" categories according to the score with the largest value. The corresponding migration matrix for the signal-like ($K=b,c,s,g$) categories is shown in Figure~\ref{fig:confusion_matrix_ii}, exhibiting small confusion across different signals. Categories containing jet pairs tagged as $K=\tau, u,d$ are discarded.

Events in each of the four $K$-like categories ($K=b,c,s,g$) are further classified, based on the value of the $K$-score, in three orthogonal categories with varying signal purity: low (LP), medium (MP) and high (HP) purity, thus leading to a total of 12 categories. The criteria to choose the thresholds on the $K$-scores that define LP, MP and HP categories were optimised by maximising the individual significance for each signal in their respective enriched category.

\section{The $jjjj$ final state}
\label{sec:jjjj}
The analysis of the $jjjj$ final state targets the $ZH$ production mode with both $Z$ and Higgs bosons decaying to pairs of partons. 
Among the three channels under study, despite the larger expected event yield, the $jjjj$ one features the worst signal-to-background ratio and presents challenges in jet clustering and resolving the ambiguity on the jet origin.

\subsection{Event selection}
\label{ssec:event_jjjj_selection}
Events containing at least one lepton ($e$ or $\mu$) with $p>20$ GeV are vetoed, for orthogonality with respect to the $\ell\ell jj$ analysis. To reject $\nu\bar{\nu}jj$ events, a selection requirement on the visible mass, $m_{\mathrm{vis}}>150\,\mathrm{GeV}$ is applied. All the visible particles are clustered in four exclusive jets with the Durham algorithm. The cosine of the angle between the visible momentum and the beam axis is required to be $\vert\cos{\theta_\mathrm{vis}}\vert<0.85$ to reject single and di-boson backgrounds. Additional requirements on the $d_{1,2}$, $d_{2,3}$ and $d_{3,4}$ merge distance variables are applied such that a four jet topology is selected.

The measurement of the reconstructed jet energies ($E_i$, $i=1..4$ ) is improved by exploiting four-momentum conservation, the known initial-state four-momentum ($\sqrt{s}, \vec{0}$), and the jet directions $\vec{\beta}_i=\vec{p}_i/E_i$, solving the following system of four equations:

\[
\begin{bmatrix}
\addlinespace
    1       & 1& 1 & 1 \\ \addlinespace
    \beta_1^x      & \beta_2^x  & \beta_3^x   & \beta_4^x   \\ \addlinespace
    \beta_1^y      & \beta_2^y  & \beta_3^y   & \beta_4^y   \\ \addlinespace
    \beta_1^z      & \beta_2^z  & \beta_3^z   & \beta_4^z   \\ \addlinespace
\end{bmatrix}
\begin{bmatrix}
 \addlinespace
E_1\\ \addlinespace
E_2\\ \addlinespace
E_3\\ \addlinespace
E_4 \\ \addlinespace
\end{bmatrix}
=
\begin{bmatrix}
 \addlinespace
\sqrt{s}\\ \addlinespace
0\\ \addlinespace
0\\ \addlinespace
0\\ \addlinespace
\end{bmatrix}
\]

The solution of the system is performed via a numerical inversion of the matrix. 
A $\chi^2$ is computed from the values of the jet energies before and after the matrix inversion, assuming a jet energy resolution of $50\%/\sqrt{E}$. Events with values of $\chi^2$ above a certain threshold are vetoed suppressing events with poor jet energy reconstruction.

The yields and efficiency of the selection after each step, for signal and background processes, are shown in Table~\ref{tab:pre_selection_jjjj}.

\begin{table}[!htbp]

  \caption{Expected event yields and efficiencies (in \%)  for signal and background processes, after each step of the $jjjj$ event selection, 
    assuming an integrated luminosity $\mathcal{L}=10.8$~\iab of $e^+e^-$ collisions at $\sqrt{s}=240\,\mathrm{GeV}$. Yields at $\sqrt{s}=365\,\mathrm{GeV}$ are shown in Table~\ref{tab:pre_selection_jjjj_365} in Appendix~\ref{appendix}. Here, $Z(q\bar{q})$ in $ZH$ processes denotes $Z$ boson decays to $u\bar{u}$ and $d\bar{d}$. Inclusive samples for $ \nu\bar{\nu}H$, $ e^+e^-H$ and $ \mu^+\mu^-H$ are included showing the orthogonality between the analyses.} \label{tab:pre_selection_jjjj} 
    \resizebox{\textwidth}{!}{

        \begin{tabular}{l|ccccc|c} \hline
					& Yield & Yield & Yield after & Yield after & Yield after&  \\ 

        Process & before & after & $m_{\mathrm{vis}},\, \theta_{\mathrm{vis}}$ & $d_{\mathrm{ij}}$ & $\chi^2$ & Efficiency (\%) \\ 
					 & selection & lepton veto & requirement &  requirement & requirement &  \\ \hline

        $ Z(c\bar{c})H(gg)$ & $2.06\times10^{4}$ & $2.06\times10^{4}$ & $2.04\times10^{4}$ & $1.81\times10^{4}$ & $1.72\times10^{4}$ & 83.5 \\
        $ Z(c\bar{c})H(s\bar{s})$ & $6.06\times10^{1}$ & $6.06\times10^{1}$ & $5.99\times10^{1}$ & $5.32\times10^{1}$ & $5.01\times10^{1}$ & 82.7 \\
        $ Z(c\bar{c})H(c\bar{c})$ & $7.29\times10^{3}$ & $7.29\times10^{3}$ & $7.04\times10^{3}$ & $6.38\times10^{3}$ & $5.89\times10^{3}$ & 80.8 \\
        $ Z(c\bar{c})H(b\bar{b})$ & $1.47\times10^{5}$ & $1.47\times10^{5}$ & $1.32\times10^{5}$ & $1.21\times10^{5}$ & $1.06\times10^{5}$ & 72.1 \\
        $ Z(q\bar{q})H(gg)$ & $4.72\times10^{4}$ & $4.72\times10^{4}$ & $4.71\times10^{4}$ & $4.05\times10^{4}$ & $3.92\times10^{4}$ & 83.1 \\
        $ Z(q\bar{q})H(s\bar{s})$ & $1.38\times10^{2}$ & $1.38\times10^{2}$ & $1.38\times10^{2}$ & $1.19\times10^{2}$ & $1.14\times10^{2}$ & 82.6 \\
        $ Z(q\bar{q})H(c\bar{c})$ & $1.67\times10^{4}$ & $1.67\times10^{4}$ & $1.63\times10^{4}$ & $1.44\times10^{4}$ & $1.36\times10^{4}$ & 81.4 \\
        $ Z(q\bar{q})H(b\bar{b})$ & $3.36\times10^{5}$ & $3.36\times10^{5}$ & $3.07\times10^{5}$ & $2.79\times10^{5}$ & $2.48\times10^{5}$ & 73.8 \\
        $ Z(b\bar{b})H(gg)$ & $2.65\times10^{4}$ & $2.65\times10^{4}$ & $2.51\times10^{4}$ & $2.29\times10^{4}$ & $2.08\times10^{4}$ & 78.5 \\
        $ Z(b\bar{b})H(s\bar{s})$ & $7.77\times10^{1}$ & $7.77\times10^{1}$ & $7.38\times10^{1}$ & $6.74\times10^{1}$ & $6.02\times10^{1}$ & 77.5 \\
        $ Z(b\bar{b})H(c\bar{c})$ & $9.36\times10^{3}$ & $9.36\times10^{3}$ & $8.61\times10^{3}$ & $7.96\times10^{3}$ & $6.98\times10^{3}$ & 74.6 \\
        $ Z(b\bar{b})H(b\bar{b})$ & $1.88\times10^{5}$ & $1.88\times10^{5}$ & $1.59\times10^{5}$ & $1.49\times10^{5}$ & $1.23\times10^{5}$ & 65.4 \\
        $ Z(s\bar{s})H(gg)$ & $2.65\times10^{4}$ & $2.65\times10^{4}$ & $2.64\times10^{4}$ & $2.29\times10^{4}$ & $2.22\times10^{4}$ & 83.8 \\
        $ Z(s\bar{s})H(s\bar{s})$ & $7.77\times10^{1}$ & $7.77\times10^{1}$ & $7.76\times10^{1}$ & $6.72\times10^{1}$ & $6.46\times10^{1}$ & 83.1 \\
        $ Z(s\bar{s})H(c\bar{c})$ & $9.35\times10^{3}$ & $9.35\times10^{3}$ & $9.15\times10^{3}$ & $8.14\times10^{3}$ & $7.67\times10^{3}$ & 82.0 \\
        $ Z(s\bar{s})H(b\bar{b})$ & $1.88\times10^{5}$ & $1.88\times10^{5}$ & $1.72\times10^{5}$ & $1.57\times10^{5}$ & $1.40\times10^{5}$ & 74.5 \\
        $ Z(b\bar{b})H(\tau\tau)$ & $2.03\times10^{4}$ & $2.03\times10^{4}$ & $1.29\times10^{4}$ & $7.34\times10^{3}$ & $2.38\times10^{3}$ & 11.7 \\
        $ Z(c\bar{c})H(\tau\tau)$ & $1.58\times10^{4}$ & $1.58\times10^{4}$ & $1.06\times10^{4}$ & $6.35\times10^{3}$ & $2.14\times10^{3}$ & 13.5 \\
        $ Z(s\bar{s})H(\tau\tau)$ & $2.03\times10^{4}$ & $2.03\times10^{4}$ & $1.38\times10^{4}$ & $8.55\times10^{3}$ & $2.88\times10^{3}$ & 14.2 \\
        $ Z(q\bar{q})H(\tau\tau)$ & $3.61\times10^{4}$ & $3.61\times10^{4}$ & $2.45\times10^{4}$ & $1.52\times10^{4}$ & $5.12\times10^{3}$ & 14.2 \\
        $ Z(b\bar{b})H(Z\gamma)$ & $4.96\times10^{2}$ & $4.96\times10^{2}$ & $4.29\times10^{2}$ & $3.06\times10^{2}$ & $2.72\times10^{2}$ & 54.8 \\
        $ Z(c\bar{c})H(Z\gamma)$ & $3.86\times10^{2}$ & $3.86\times10^{2}$ & $3.50\times10^{2}$ & $2.45\times10^{2}$ & $2.27\times10^{2}$ & 58.8 \\
        $ Z(s\bar{s})H(Z\gamma)$ & $4.96\times10^{2}$ & $4.96\times10^{2}$ & $4.55\times10^{2}$ & $3.12\times10^{2}$ & $2.93\times10^{2}$ & 59.1 \\
        $ Z(q\bar{q})H(Z\gamma)$ & $8.68\times10^{2}$ & $8.68\times10^{2}$ & $7.95\times10^{2}$ & $5.42\times10^{2}$ & $5.09\times10^{2}$ & 58.6 \\
        $ Z(b\bar{b})H(WW)$ & $6.97\times10^{4}$ & $6.97\times10^{4}$ & $4.51\times10^{4}$ & $3.83\times10^{4}$ & $2.73\times10^{4}$ & 39.2 \\
        $ Z(c\bar{c})H(WW)$ & $5.42\times10^{4}$ & $5.42\times10^{4}$ & $3.69\times10^{4}$ & $3.11\times10^{4}$ & $2.28\times10^{4}$ & 42.1 \\
        $ Z(s\bar{s})H(WW)$ & $6.96\times10^{4}$ & $6.96\times10^{4}$ & $4.79\times10^{4}$ & $4.00\times10^{4}$ & $2.95\times10^{4}$ & 42.4 \\
        $ Z(q\bar{q})H(WW)$ & $1.24\times10^{5}$ & $1.24\times10^{5}$ & $8.51\times10^{4}$ & $7.07\times10^{4}$ & $5.21\times10^{4}$ & 42.0 \\
        $ Z(b\bar{b})H(ZZ)$ & $8.55\times10^{3}$ & $8.55\times10^{3}$ & $6.99\times10^{3}$ & $4.99\times10^{3}$ & $3.60\times10^{3}$ & 42.1 \\
        $ Z(c\bar{c})H(ZZ)$ & $6.66\times10^{3}$ & $6.66\times10^{3}$ & $5.73\times10^{3}$ & $4.07\times10^{3}$ & $3.04\times10^{3}$ & 45.6 \\
        $ Z(s\bar{s})H(ZZ)$ & $8.54\times10^{3}$ & $8.54\times10^{3}$ & $7.48\times10^{3}$ & $5.29\times10^{3}$ & $3.99\times10^{3}$ & 46.7 \\
        $ Z(q\bar{q})H(ZZ)$ & $1.52\times10^{4}$ & $1.52\times10^{4}$ & $1.33\times10^{4}$ & $9.36\times10^{3}$ & $7.06\times10^{3}$ & 46.4 \\
        $ W^+W^-$ & $1.64\times10^{8}$ & $1.64\times10^{8}$ & $1.05\times10^{8}$ & $6.92\times10^{7}$ & $5.85\times10^{7}$ & 35.7 \\
        $ ZZ$ & $1.36\times10^{7}$ & $1.36\times10^{7}$ & $1.09\times10^{7}$ & $5.69\times10^{6}$ & $4.69\times10^{6}$ & 34.5 \\
        $ q\bar{q}$ & $5.65\times10^{8}$ & $5.65\times10^{8}$ & $5.45\times10^{8}$ & $1.04\times10^{8}$ & $4.02\times10^{7}$ & 7.1 \\
        $ e^+e^-H$ & $7.66\times10^{4}$ & $7.66\times10^{4}$ & $1.81\times10^{2}$ & $1.13\times10^{2}$ & $0.0$ & 0.0 \\
        $ \mu^+\mu^-H$ & $7.21\times10^{4}$ & $7.21\times10^{4}$ & $9.42\times10^{1}$ & $3.68\times10^{1}$ & $0.0$ & 0.0 \\
        $ \nu\bar{\nu}H$ & $4.71\times10^{5}$ & $4.71\times10^{5}$ & $4.12\times10^{5}$ & $2.29\times10^{2}$ & $0.0$ & $0.0$ \\
        \hline
    \end{tabular}
    }
\end{table}

\subsection{Event classification}
After the previous preselection, the four jets are combined pair-wise to form the Higgs and $Z$ boson candidates. The procedure, illustrated in Figure~\ref{fig:pairingFlowChart}, proceeds as follows:

\begin{figure}[!htbp]
\centering
\includegraphics[width=0.99\textwidth]{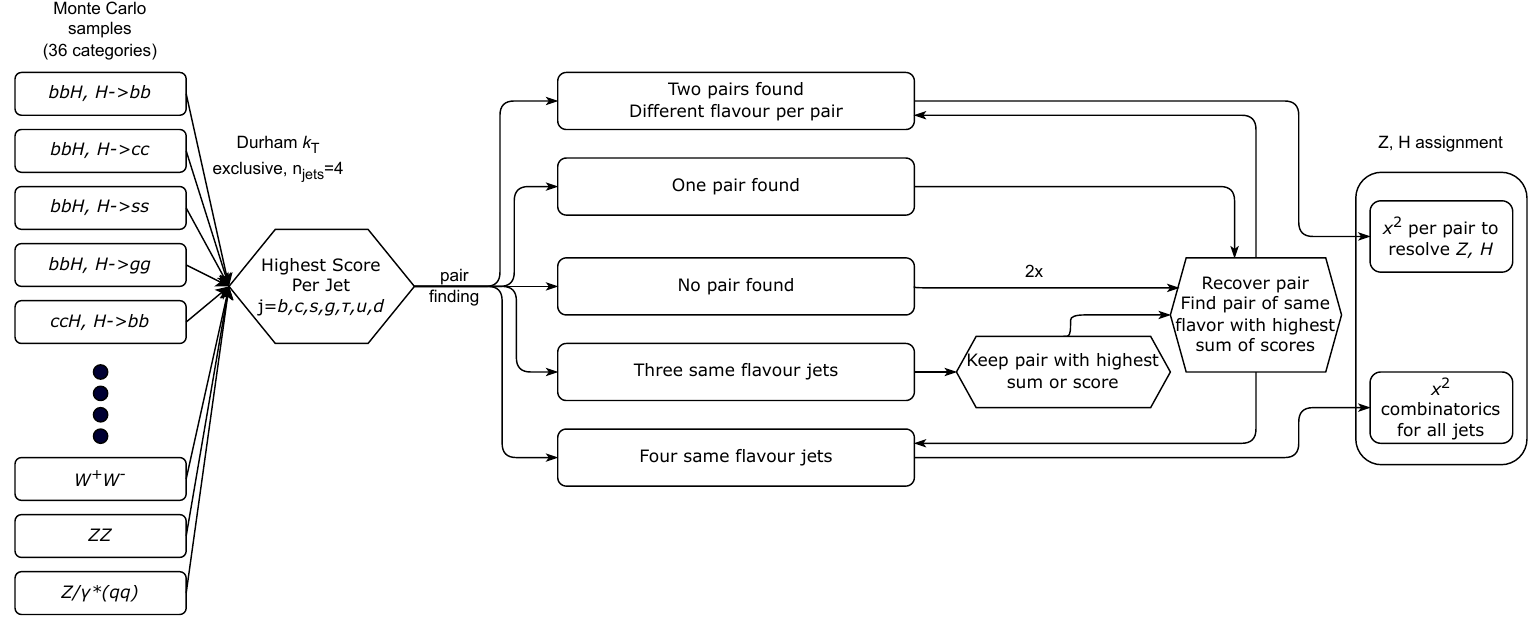}
\caption{Illustration of the jet pairing procedure in the fully-hadronic analysis.}
\label{fig:pairingFlowChart}
\end{figure}

\begin{itemize}
\item 
First, the flavour of the four jets is determined.
Each reconstructed jet $i$ is initially assigned a flavour corresponding to the highest $K_i$-score from the flavour tagging algorithm ($K=b,c,s,g,\tau, u, d$).
If the event does not contain two pairs of jets with the same predicted flavour, a recovery procedure is performed:

\begin{enumerate}[label=(\alph*)]
\item If there is one pair of jets with the same flavour, while the remaining two have different flavours (e.g. $b,b,g,c$), the sum of the $K$-scores of the two different-flavour jets is computed for each hypothesis $K=b,c,s,g,\tau, u, d$. The two jets are then assigned the flavour hypothesis that maximises this sum.
\item If no pair of jets shares the same flavour  (e.g. $b, g, c, s$), the procedure above is applied twice: first to identify the jet pair and flavour hypothesis that yield the highest total $K$-score among all possible pairings, and then again to the remaining pair of jets.
\item  If three jets share the same flavour (e.g. $b, b, b, c$), the pair among the three identical-flavour jets with the highest combined $K$-score for that flavour hypothesis is retained.
The remaining pair of jets is then assigned the flavour hypothesis that maximises the sum of their $K$-scores.
\end{enumerate}

\item 
After the previous steps, the events contain two pairs of same-flavoured jet.
The assignment of each jet either a parent $Z$ or Higgs boson depends on whether the pairs are of same or different flavour:
\begin{enumerate}[label=(\alph*)]
\item if the pairs have different flavour, then one pair is considered to originate from the $Z$ boson and the other one from the Higgs boson, choosing the configuration that minimises $\chi^2= (m_{j_1 j_2}-m_Z)^2+(m_{j_3 j_4}-m_H)^2$, where $m_{j_1 j_2},~m_{j_3 j_4}$, are the di-jet invariant masses and $m_Z, m_H$ are the $Z$ and Higgs boson masses ($m_Z=91.2\,\mathrm{GeV}~\textrm{and}~ m_H=125\,\mathrm{GeV}$).
\item if all four jets are of the same flavour, then  all possible pairs of jets are considered, and the configuration that minimises $\chi^2= (m_{j_1 j_2}-m_Z)^2+(m_{j_3 j_4}-m_H)^2$ is chosen.
\end{enumerate}
\end{itemize}

\begin{figure}[!htbp]
  \centering
    \includegraphics[width=0.49\textwidth]{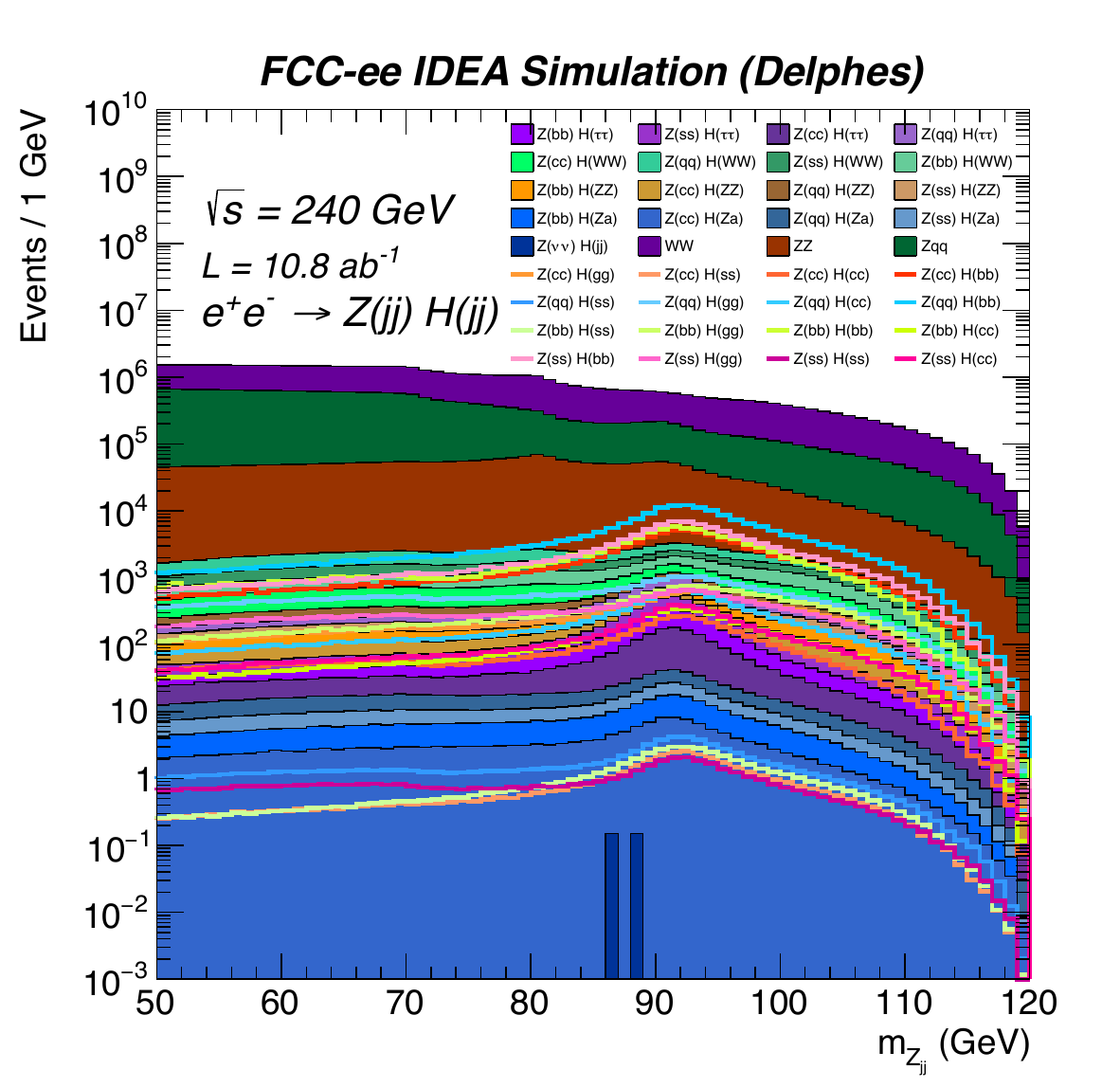}
   \includegraphics[width=0.49\textwidth]{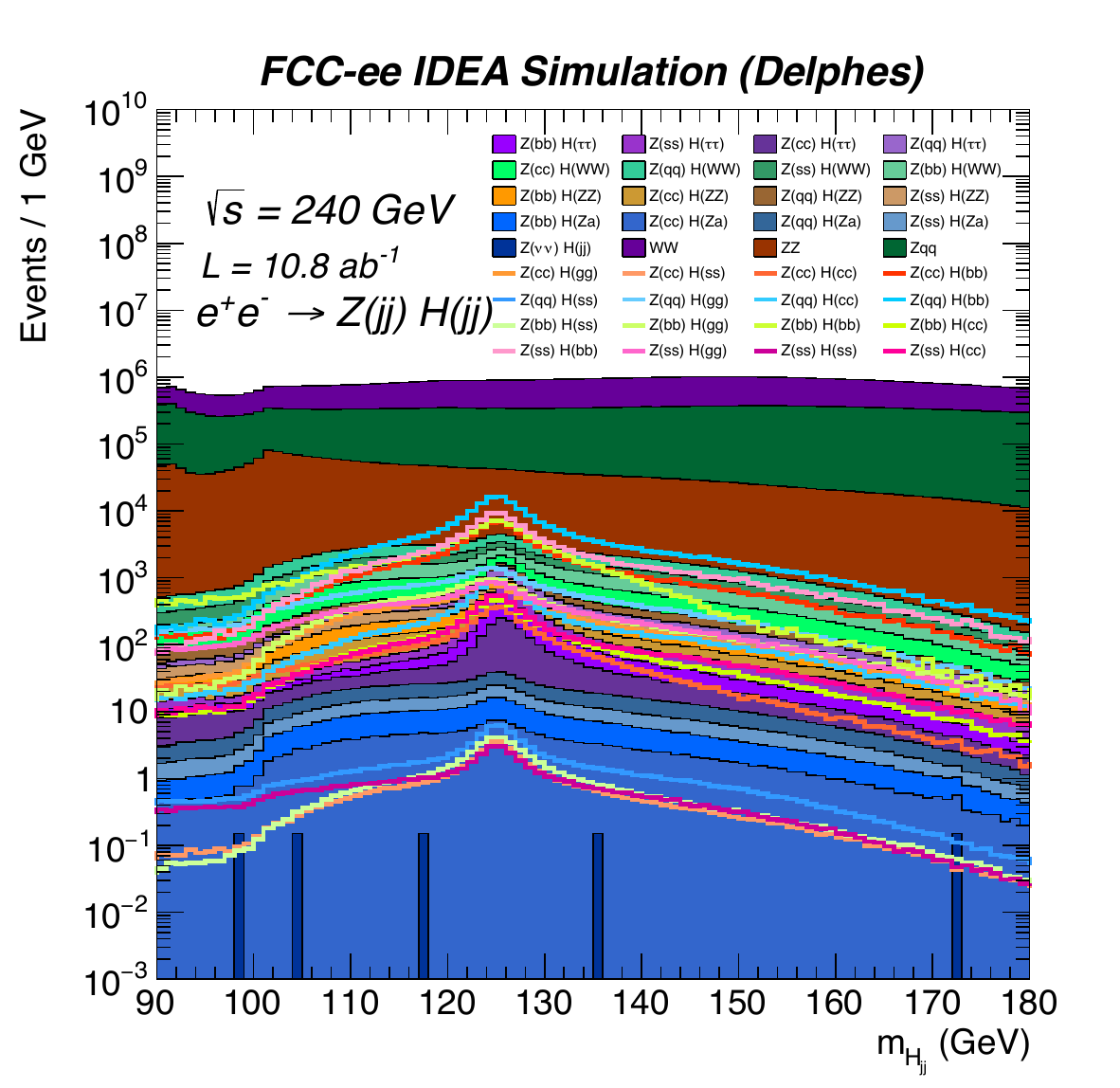}
      \includegraphics[width=0.49\textwidth]{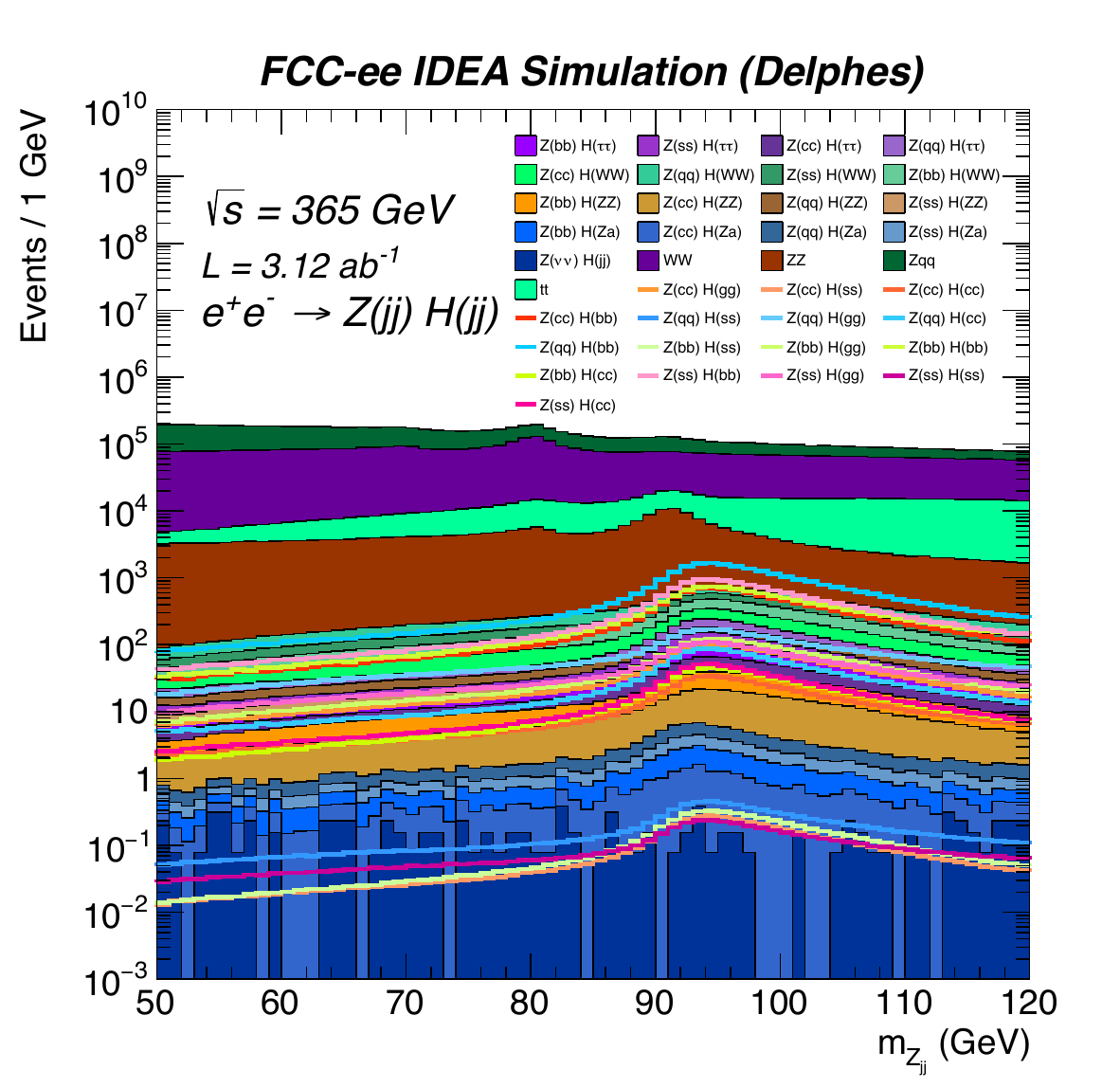}
   \includegraphics[width=0.49\textwidth]{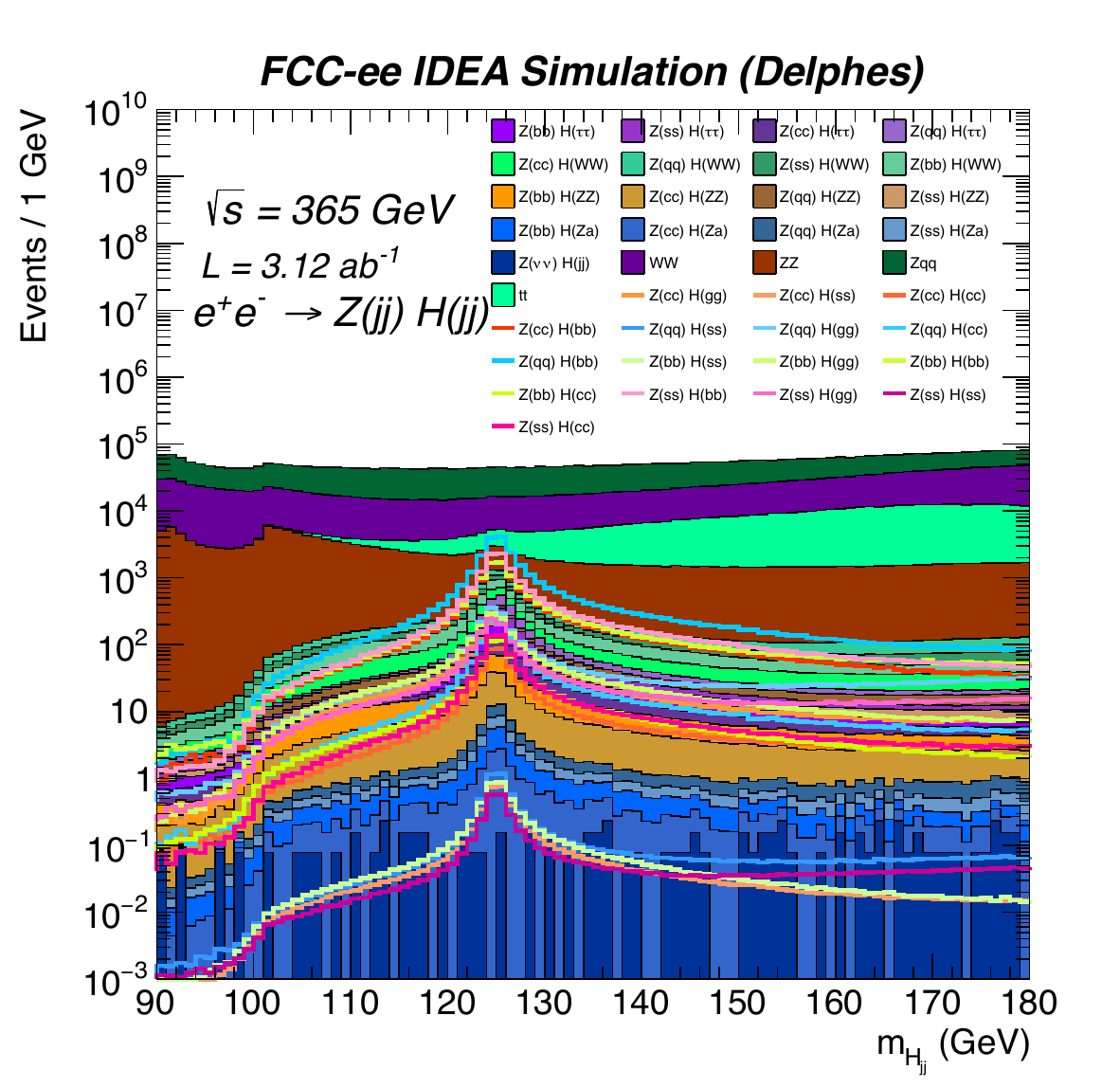}
  \caption{Distribution of di-jet invariant mass $m_{jj}$ of the jets associated to the $Z$ (left) and to the $H$ (right) bosons for $\sqrt{s}=240\,\mathrm{GeV}$ (top) and $\sqrt{s}=365\,\mathrm{GeV}$ (bottom).
  }
  \label{fig:mz_mh_fullhadronic}
\end{figure}

At the end of the procedure, the four jets are thus grouped into two pairs of jets of same flavour, considered to be candidates $Z$ ($Z_{jj}$) or Higgs ($H_{jj}$) boson decays.
To further suppress $WW$ and $ZZ$ background events, the following requirements are applied:
\begin{eqnarray}
   \sqrt{(m_{Z_{jj}}-m_W)^2+(m_{H_{jj}} - m_W)^2}>10~\mathrm{GeV}, \\
   \sqrt{(m_{Z_{jj}}-m_Z)^2+(m_{H_{jj}}-m_Z)^2}>10~\mathrm{GeV},
\end{eqnarray}
where $m_W=80.377\,\mathrm{GeV}$ is the $W$ boson mass. 
Finally, $m_{Z_{jj}}$ and $ m_{H_{jj}}$ are required to be in the following ranges: $50~\mathrm{GeV}<m_{Z_{jj}}<125~\mathrm{GeV}$, and $91~\mathrm{GeV}<m_{H_{jj}}<180~\mathrm{GeV}$. 
The $m_{Z_{jj}}$ and $m_{H_{jj}}$ distributions of signal and background events passing the selection are shown in Figure~\ref{fig:mz_mh_fullhadronic}.

\begin{figure}[!htbp]
  \begin{center}
  \includegraphics[width=0.99\textwidth]{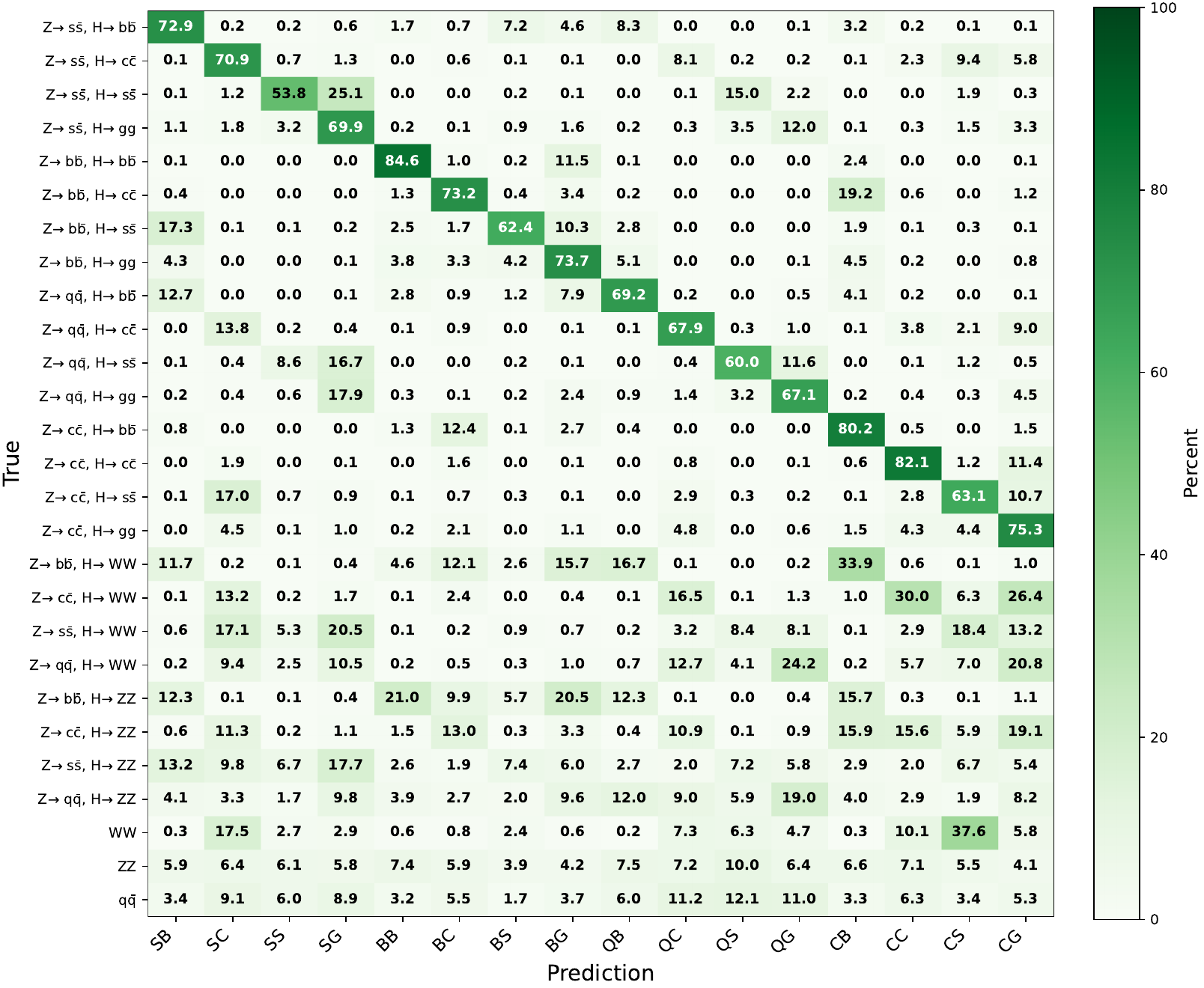}
  \caption{Migration matrix of the full hadronic final state. Each row corresponds to a different process. The 16 columns show the fractions of events in these jet combinations entering the fit. For example SB would include events tagged as $Z\to s\bar{s},~ H\to b\bar{b}$. Events where Higgs boson decays are tagged as $\tau\tau$ are removed.}
  \label{fig:confusion_matrix_jjjj}
  \end{center}
  \end{figure}

Events passing the previous requirements are classified in 16 orthogonal regions ($BB$, $BC$, ..) based on the flavours assigned to the two jet pairs. Events in which the Higgs boson decay products are tagged as $\tau$ leptons are rejected, suppressing more than $95\%$ of $H\to \tau\tau$ events.
The corresponding migration matrix is shown in Figure~\ref{fig:confusion_matrix_jjjj}.

The events in each of the 16 regions are further classified into three orthogonal categories of low (LP), medium (MP) and high (HP) signal purity based on selections on the sum of the $K$-scores for the two jet pairs, leading to a total of 48 categories.
The boundaries that define LP, MP and HP categories were chosen by maximising the individual significance for each signal in their respective enriched category.

\section{Signal strength measurement}
\label{sec:globalfit}

One- or two-dimensional distributions of discriminant variables in pseudo-data samples at $\sqrt{s}=240$ or $365\,\mathrm{GeV}$ in the categories of the three different channels are interpreted simultaneously in terms of signal strengths of the Higgs boson production and decay processes under study through a binned maximum likelihood fit.

Pseudo-data samples are obtained by summing together the histograms of the discriminant variables for each signal and background process, scaled to the assumed value of the integrated luminosity given the SM cross-sections and branching ratios of each process.

In each category, the likelihood function is constructed as the product of Poisson probabilities over all bins of the histogram of the corresponding discriminant variable(s), where the observed count in the pseudo-data sample is compared to the expected yield from the sum of signal and background. The likelihood function at each centre-of-mass energy is calculated as the product of the likelihood functions for the categories of the three final states under study.

The yields for each Higgs boson decay mode in each bin are calculated as the product of the expected yield times a signal strength $\mu_{XX}$.
Separate signal strengths are assigned to each Higgs boson decay mode $XX$ under study: $\bb$, $\cc$, $s\bar{s}$, $gg$. 
The large control regions (categories targeting background processes or Higgs boson decays other than those under study) and the sidebands of the discriminant variable distributions in the categories targeting the signal processes allow fitting also the 
normalisations of other Higgs boson decays ($\tau\tau$, $WW$ and $ZZ$) and of the main background processes. 
Tests showed that, allowing them to float freely or constraining them with a prior at the few-percent level, yields consistent results and hence are left floating.

The discriminant variables used in the fit in each channel are the following:
\begin{itemize}
\item $\ell\ell jj$ final state: one-dimensional distributions of the recoil mass ($m_{\textrm{recoil}}$) in the interval 120--140$\,\textrm{GeV}$, with a bin width of  0.25$\,\textrm{GeV}$, are used.
\item $\nu\bar{\nu}jj$ final state: two-dimensional distributions of the recoil mass ($m_{\textrm{recoil}}$) versus the di-jet invariant mass  ($m_{\textrm{jj}}$) are used. The ranges for $m_{\textrm{jj}}$ and $m_{\textrm{rec}}$ are $50-180\,\mathrm{GeV}$ ($50-180$) and $20-200\,\mathrm{GeV}$ ($20-280$) for $\sqrt{s}=240$~(365)~GeV, with bin widths of $1\,\mathrm{GeV}$ and $5\,\mathrm{GeV}$ respectively.

\item $jjjj$ final state: two-dimensional distributions of the di-jet invariant masses   $m_{H_{jj}}$ and $m_{Z_{jj}}$ are used, in the ranges  of $90-180\,\mathrm{GeV}$ and $50-120\,\mathrm{GeV}$, with bin widths of $1\,\mathrm{GeV}$ and $5\,\mathrm{GeV}$ respectively.
\end{itemize}

In the $\nu\bar{\nu} jj$ final state, separate  distributions of $m_{\textrm{jj}}$ vs $m_{\textrm{recoil}}$ and signal strengths are included in the fit for the $ZH$ and VBF production modes, exploiting the expected differences in these distributions for the two production modes to distinguish among them.

Uncertainties from the finite size of simulated samples are incorporated in the fit using the Beeston–Barlow method~\cite{Barlow:1993dm}. Shape systematic uncertainties from background modelling, as well as experimental uncertainties related to event selection and flavour tagging, are not included. Large control samples at FCC-ee and improvements in theoretical calculations in the next two decades will allow the shapes and normalisations of these processes to be constrained with an accuracy at the permille level.

The uncertainties in the signal strengths for the Higgs boson decay modes under study and the other Higgs boson decays expected from each individual channel, when inputs from the other channels are not included in the fit, are summarised in Table~\ref{tab:fitresultsIndividual_zh_vbf}.
The uncertainties expected when the three channels are combined together are shown in Table~\ref{tab:fitresultsGlobal_zh_vbf}.

Signal strengths for the Higgs boson decay modes that are backgrounds to those under study are constrained to precisions that are at best 1.1\% ($H\to WW$), 2.5\% ($\tau\tau$) and 6.9\% ($ZZ$) for $ZH$ production at $\sqrt{s}=240$~GeV. Dedicated measurements targeting these processes as signals will achieve significantly better precision.

\begin{table}[ht]
    \centering
    \caption{Relative uncertainty (\%) for  $\sigma_{ZH}\times \mathcal{B}(H\rightarrow XX)$ and  $\sigma_{\nu_e\bar{\nu}_e}\times \mathcal{B}(H\rightarrow XX)$ at $68\,\%$ CL for the individual $\ellell jj$, $\nu\nu jj$, and $jjjj$ final states, assuming integrated luminosities of $\mathcal{L}$=10.8 ab$^{-1}$ at $\sqrt{s}=240\,$GeV and  $\mathcal{L}$=3.12 ab$^{-1}$ at $\sqrt{s}=365\,$GeV.}

    \renewcommand{\arraystretch}{1.3}
    \begin{tabular}{l|cc|cc|cc|cc}
        \toprule
        Final State &\multicolumn{2}{c|}{$\ell\ell jj$} & \multicolumn{4}{c|}{$\nu\bar{\nu}jj$} & \multicolumn{2}{c}{$jjjj$}\\        
        $\sqrt{s}$~(GeV) & $240$ & $365$ & \multicolumn{2}{c|}{$240$} & \multicolumn{2}{c|}{$365$} &$240$ &$365$ \\
        Channel &\multicolumn{2}{c|}{$ZH$}  & $ZH$ & $\nu_e \bar{\nu}_e H$ & $ZH$ & $\nu_e \bar{\nu}_e H$  &\multicolumn{2}{c}{$ZH$} \\
        \midrule
        $H\to b\bar{b}$      & $\pm 0.60$ & $\pm 1.30 $  & $\pm 0.38$   & $\pm 1.99$  & $\pm 0.70$   & $\pm 0.65$    &$\pm 0.32$    &$\pm 0.54$     \\ 
        $H\to c\bar{c}$      & $\pm 3.47$ & $\pm 9.3 $  & $\pm 2.66$   & $\pm 21$     & $\pm 4.72$   & $\pm 3.42$    &$\pm 3.52$    &$\pm 4.59$     \\
	    $H\to s\bar{s}$      & $\pm 220 $ & $\pm 880  $  & $\pm 159 $   & $\pm 1000$         & $\pm 460 $   & $\pm 283$     &$\pm 480$     &$\pm 690$      \\
        $H\to gg$            & $\pm 1.93$ & $\pm 4.43 $  & $\pm 1.06$   & $\pm 5.7$   & $\pm 2.89$   & $\pm 2.58$    &$\pm 3.07$    &$\pm  4.28$    \\
        \bottomrule
    \end{tabular}
        \label{tab:fitresultsIndividual_zh_vbf}
\end{table}

\begin{table}[ht]
    \centering
    \caption{Relative uncertainty (\%) for  $\sigma_{ZH}\times \mathcal{B}(H\rightarrow XX)$ and  $\sigma_{\nu_e\bar{\nu}_e}\times \mathcal{B}(H\rightarrow XX)$ at $68\,\%$ CL combining the $\ellell jj$, $\nu\nu jj$, and $jjjj$ final states.}

    \renewcommand{\arraystretch}{1.3}

    \begin{tabular}{l | c c | c c}
        \toprule
        $\sqrt{s}$ & \multicolumn{2}{c|}{$240\,$GeV} & \multicolumn{2}{c}{$365\,$GeV} \\
        $\mathcal{L}$ & \multicolumn{2}{c|}{10.8 ab$^{-1}$} & \multicolumn{2}{c}{3.12 ab$^{-1}$} \\
        Channel & $ZH$ & $\nu_e \bar{\nu}_e H$ & $ZH$ & $\nu_e \bar{\nu}_e H$ \\
        \midrule
        $H\to b\bar{b}$      & $\pm 0.21$  & $\pm 1.89$  & $\pm 0.39$  & $\pm 0.64$ \\ 
        $H\to c\bar{c}$      & $\pm 1.75$  & $\pm 20$    & $\pm 3.01$  & $\pm 3.36$ \\
	    $H\to s\bar{s}$      & $\pm 110 $  & $\pm 990$   & $\pm 340$   & $\pm 280$ \\
        $H\to gg$            & $\pm 0.85$  & $\pm 5.5$  & $\pm 2.13$  & $\pm 2.56$ \\
        \bottomrule
    \end{tabular}

        \label{tab:fitresultsGlobal_zh_vbf}
\end{table}

\FloatBarrier

\section{Conclusion}

A study of the sensitivity to the measurement of the signal
strengths for Higgs boson decays to \bb, \cc, $gg$ and \ssbar
in $ZH$ and VBF events using $\ell\ell jj$ ($\ell=e,\mu$),
$\nunu jj$ and $jjjj$ final states at FCC-ee has been presented. The results assume an integrated luminosity of $10.8\,\mathrm{ab}^{-1}$
at $\sqrt{s}=240\,\mathrm{GeV}$ and $3.12\,\mathrm{ab}^{-1}$ at $\sqrt{s}=365\,\mathrm{GeV}$ of \epem collisions, reconstructed by assuming four 
detectors with performance identical to that of the IDEA detector concept.

The combination of the three final states leads to expected
uncertainties in the $ZH$ production signal strength 
of 0.21\%, 1.75\%, 0.85\% and 110\% for \hbb, \hcc, \hgg
and \hssbar\ decays at $\sqrt{s}=240\,\mathrm{GeV}$, while the
uncertainties are 2--3 times worse at $365\,\mathrm{GeV}$ due to the 
lower integrated luminosity and signal cross-section. A significant contribution to the $\nu \bar{\nu} H$ final state,
especially at $\sqrt{s}=365$~GeV, is expected from $W$-boson
fusion events. A dedicated analysis that measures separately the $Z(\nunu)H$ and $\nu_e\bar{\nu_e}H$ contributions for the first time yields an expected precision on the $\nu_e\bar{\nu_e}H$ signal strength below one \% for \hbb and few-\% level for \hcc and \hgg\ at $\sqrt{s}=365\,\mathrm{GeV}$.

This work constitutes the first sensitivity study of all major hadronic Higgs decay modes in a combined fit at the FCC-ee, simultaneously accounting for both $ZH$ and VBF production processes, including a full treatment of interference effects in the $\nu\bar{\nu} jj$ final state. The combination of results at $\sqrt{s}=240$ and $365\,\mathrm{GeV}$, accounting for the correlations across all production and decay modes, provides the expected FCC-ee sensitivity to hadronic Higgs decays, delivering percent to per-mille level precision for the dominant channels. Together with the precise determination of the total Higgs production cross-section from the recoil-mass method~\cite{eysermans_2025_jfb44-s0d81}, these results provide an essential input to model-independent Higgs coupling fits and to the extraction of the total Higgs width at the FCC-ee~\cite{fsrPED}.

For the first time, the sensitivity to the rare decay $H\to s\bar{s}$ is investigated in a global fit together with the other hadronic modes, demonstrating that FCC-ee could potentially provide evidence for the strange-quark Yukawa coupling. The sensitivity to the rare $H\to s\bar{s}$ decay highlights the importance of equipping the FCC-ee detector with highly performant charged hadron particle identification capabilities, as discussed in Section~\ref{sec:detector}. The ability to achieve high momentum $K/\pi$ separation up to momenta of $30\,\mathrm{GeV}$, through ionisation energy-loss measurements in a gaseous tracker or via a dedicated Cherenkov-based detector, is a key ingredient to enable efficient strange jet tagging. Dedicated detector performance studies have demonstrated that such performance is essential to reach competitive sensitivity to the strange-quark Yukawa coupling~\cite{fsrPED}.

\appendix
\section{Appendix}
\label{appendix}
\begin{figure}[!htbp]
  \centering

  \subfloat[]{\includegraphics[width=0.48\textwidth]{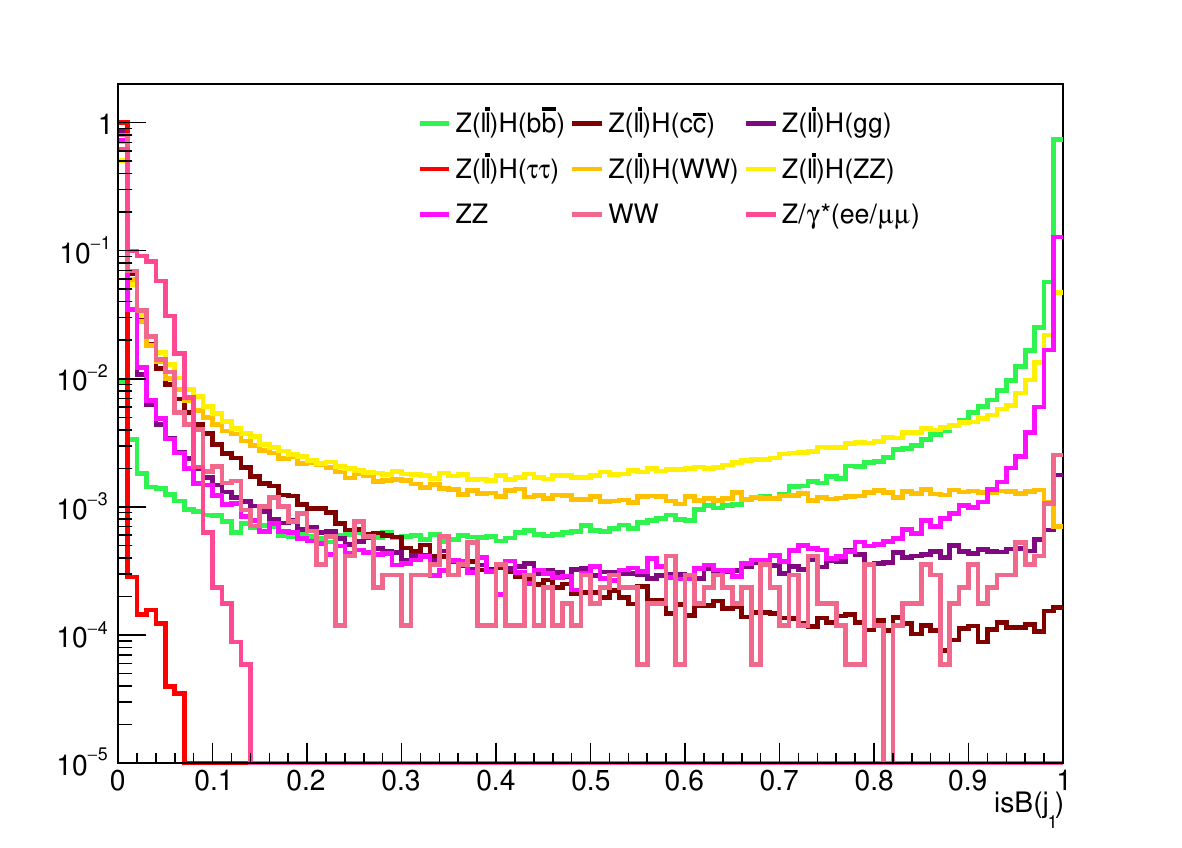}}
  \subfloat[]{\includegraphics[width=0.48\textwidth]{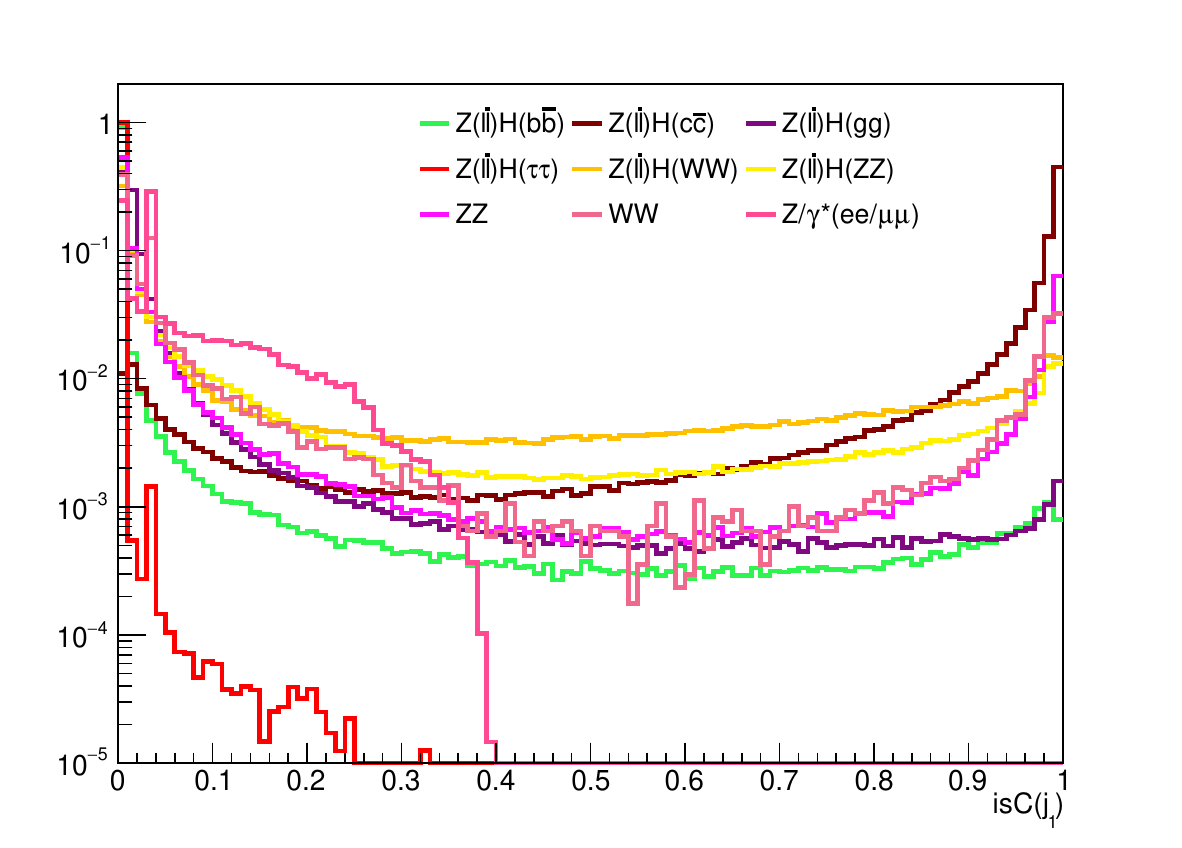}} \\

  \subfloat[]{\includegraphics[width=0.48\textwidth]{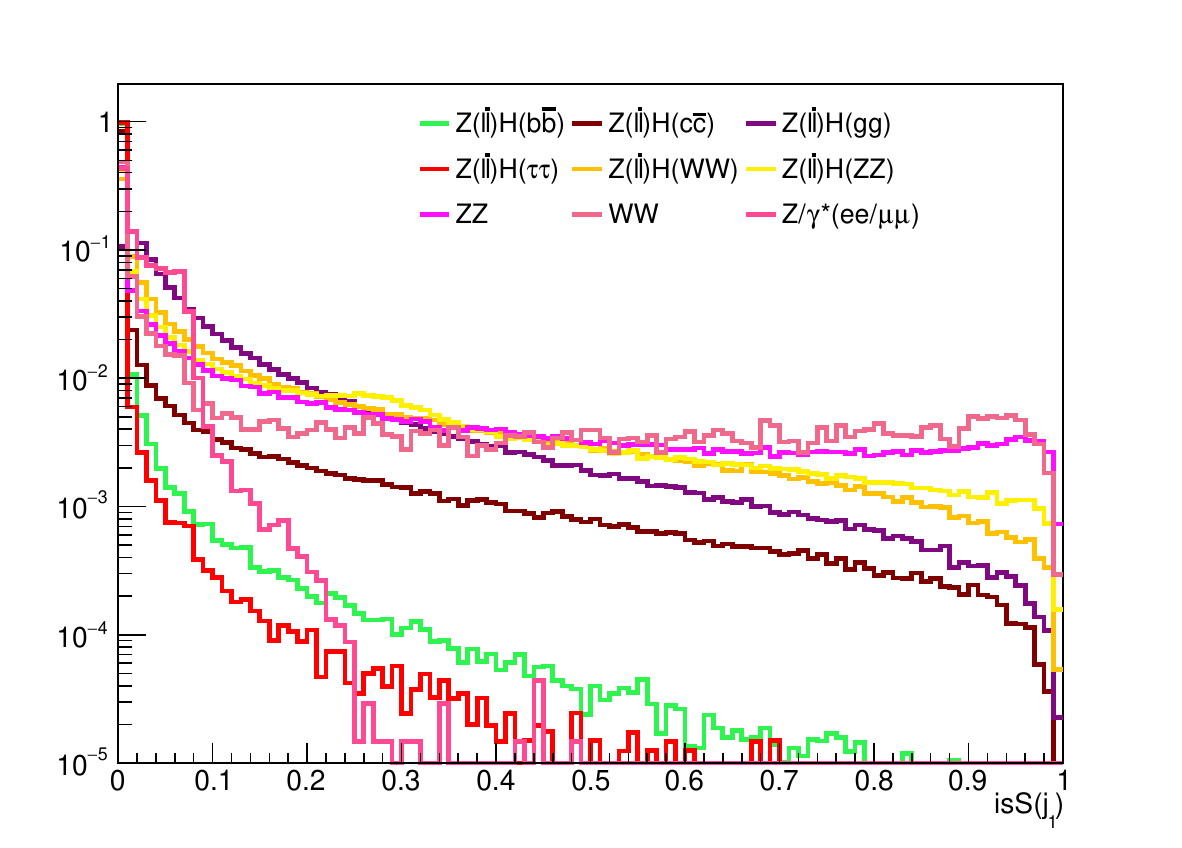}}
  \subfloat[]{\includegraphics[width=0.48\textwidth]{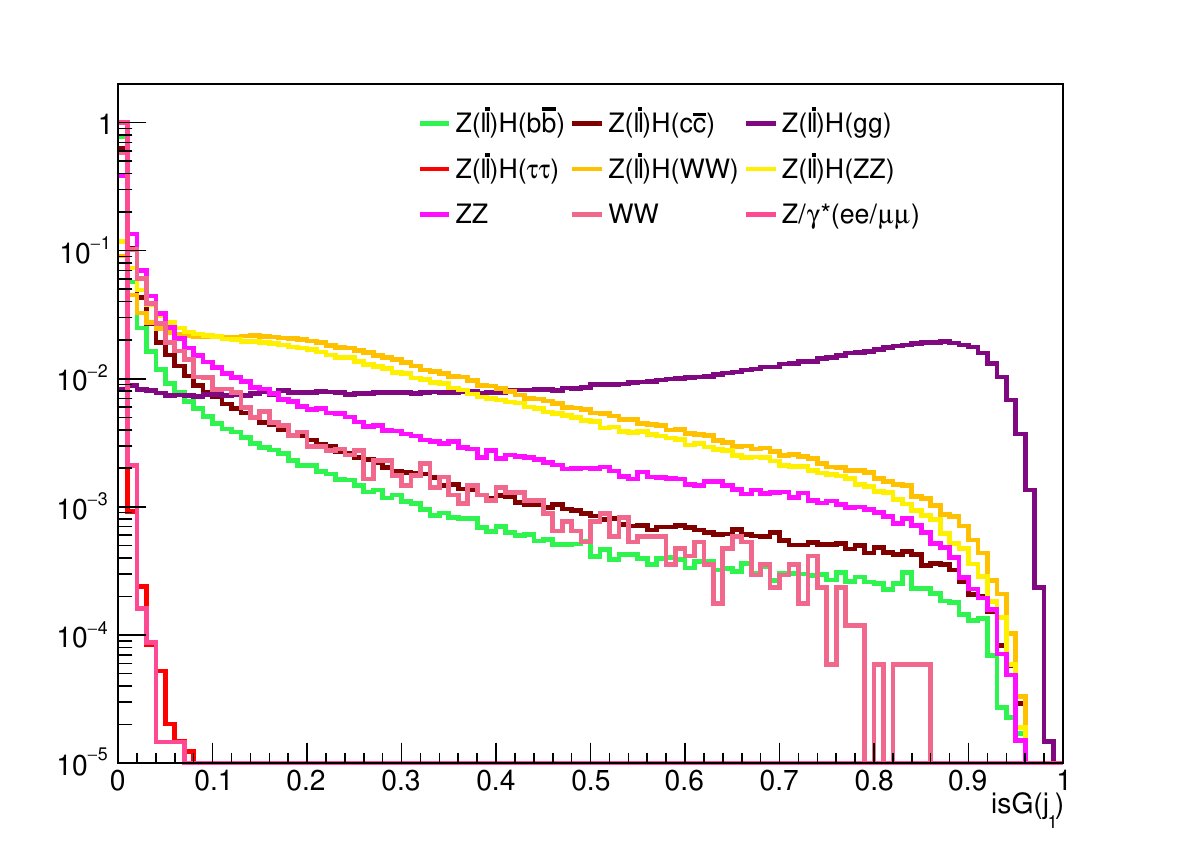}} \\

  \caption{Distributions (normalised to unit area) in $\ell\ell H$
    simulated events for selected input variables used in the neural network
    classification algorithm in the $\ell\ell jj$ analysis, namely the $b$, $c$, $s$ and $g$ tagging scores
    for the leading jet.}
  \label{fig:llh_nn_input1}
\end{figure}

\begin{figure}[!htbp]%
  \centering
  \subfloat[]{\includegraphics[width=0.48\textwidth]{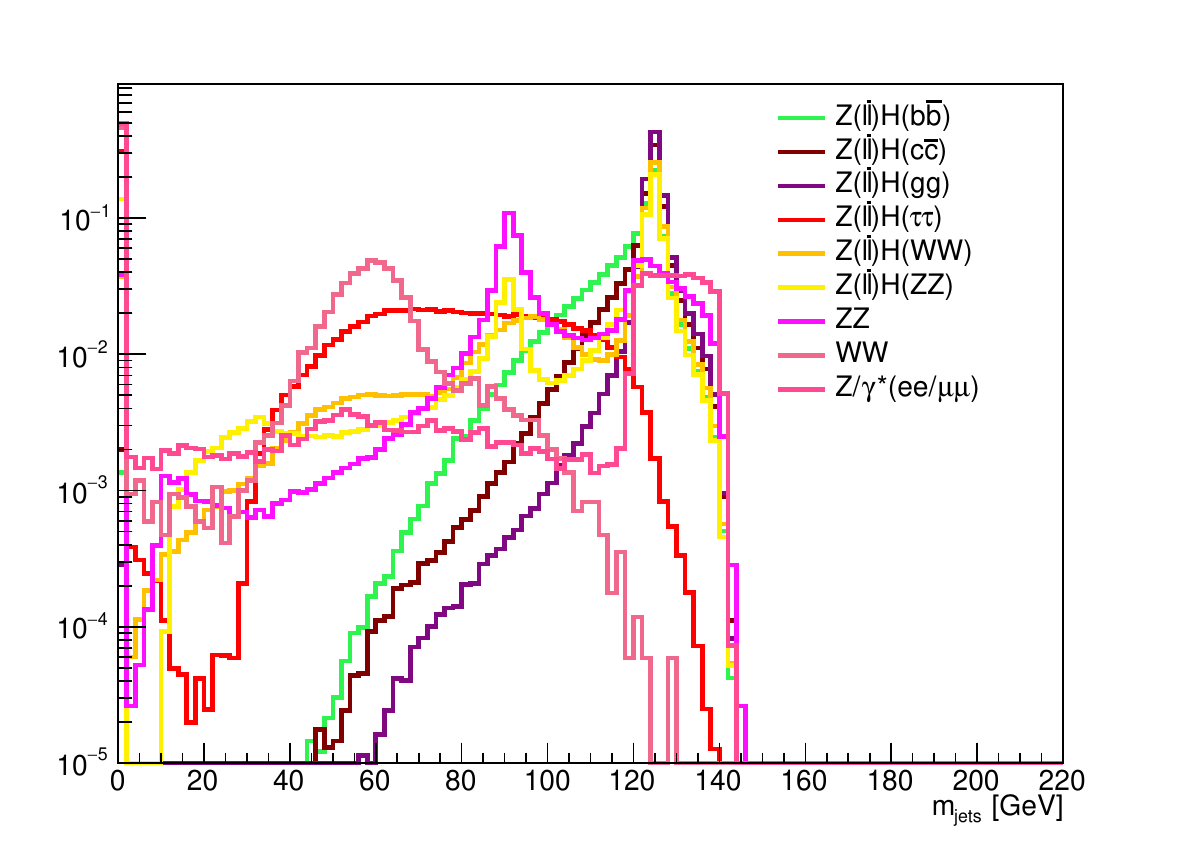}}
  \subfloat[]{\includegraphics[width=0.48\textwidth]{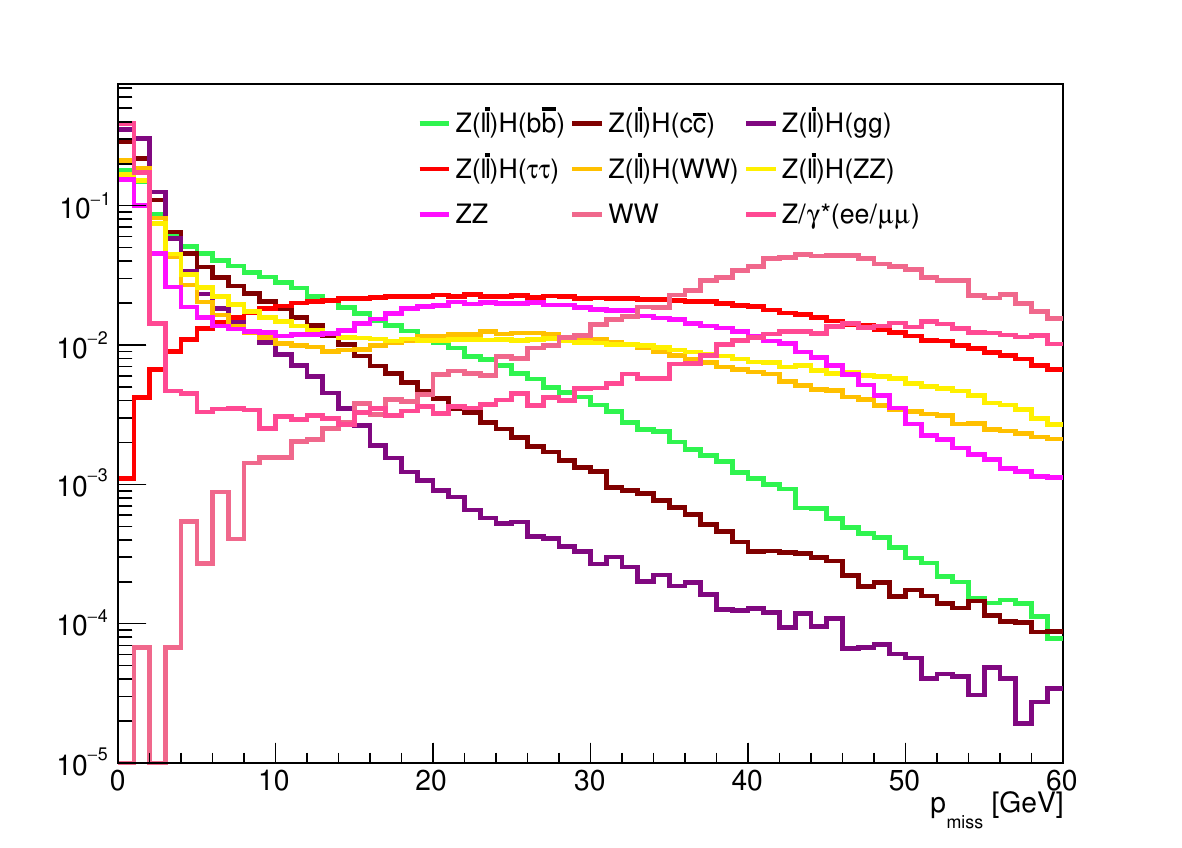}}  \\
  \subfloat[]{\includegraphics[width=0.48\textwidth]{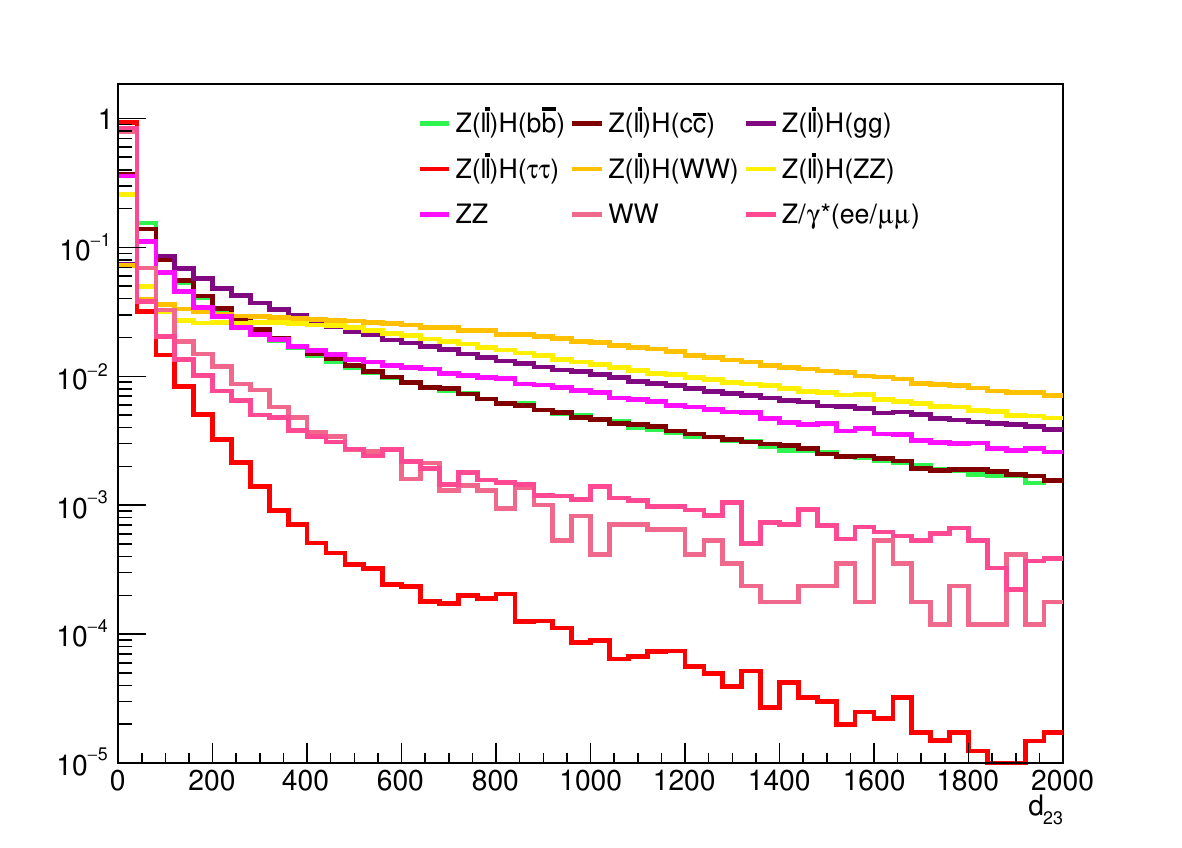}}
  \subfloat[]{\includegraphics[width=0.48\textwidth]{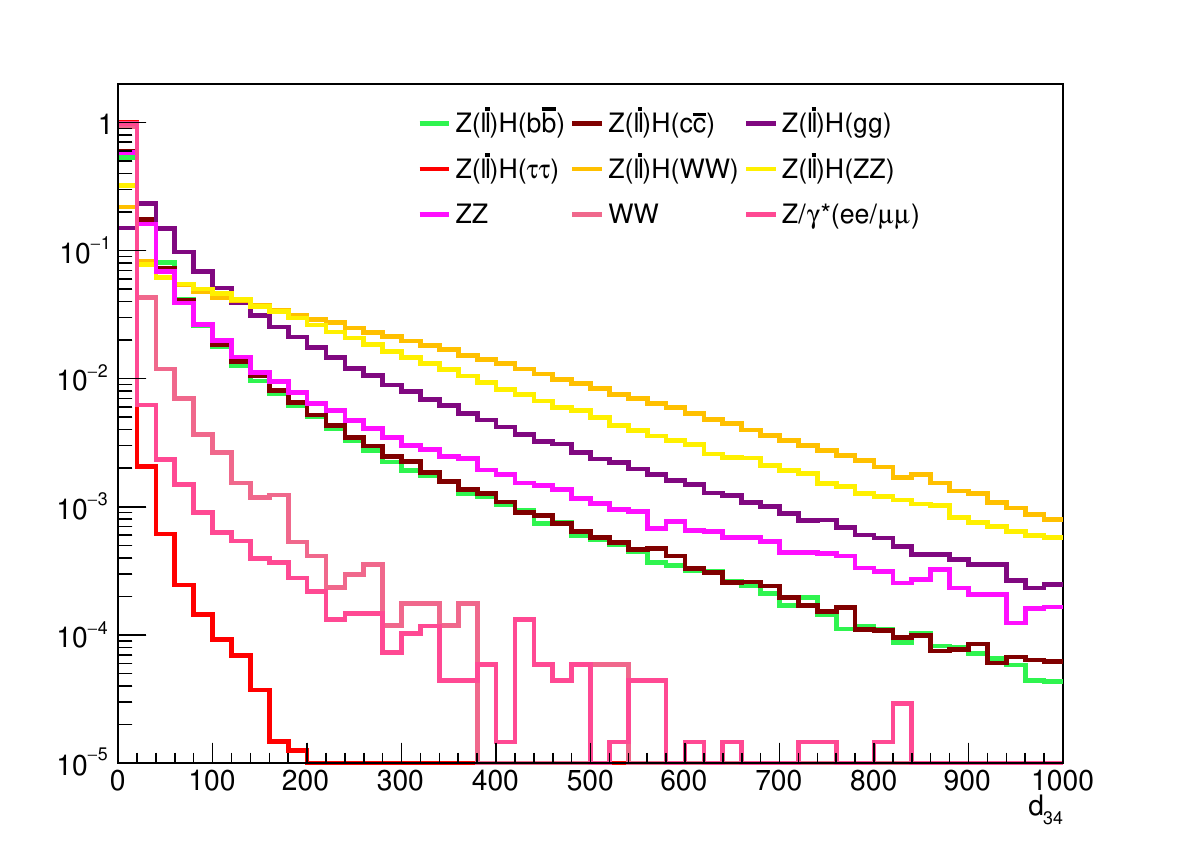}}  \\    
  \caption{Distributions (normalised to unit area) in $\ell\ell H$ simulated
    events for selected input variables used in the neural network classification
    algorithm in the $\ell\ell jj$ analysis, namely the di-jet invariant mass, the missing momentum, and the
    jet clustering distance variables $d_{23}$ and $d_{34}$.}
\label{fig:llh_nn_input3}
\end{figure}

The distributions of selected NN input variables for the $\ell\ell jj$ analysis are shown in
Figures~\ref{fig:llh_nn_input1}--\ref{fig:llh_nn_input3}.

\begin{figure}[!htbp]
  \centering
    \includegraphics[width=0.49\textwidth]{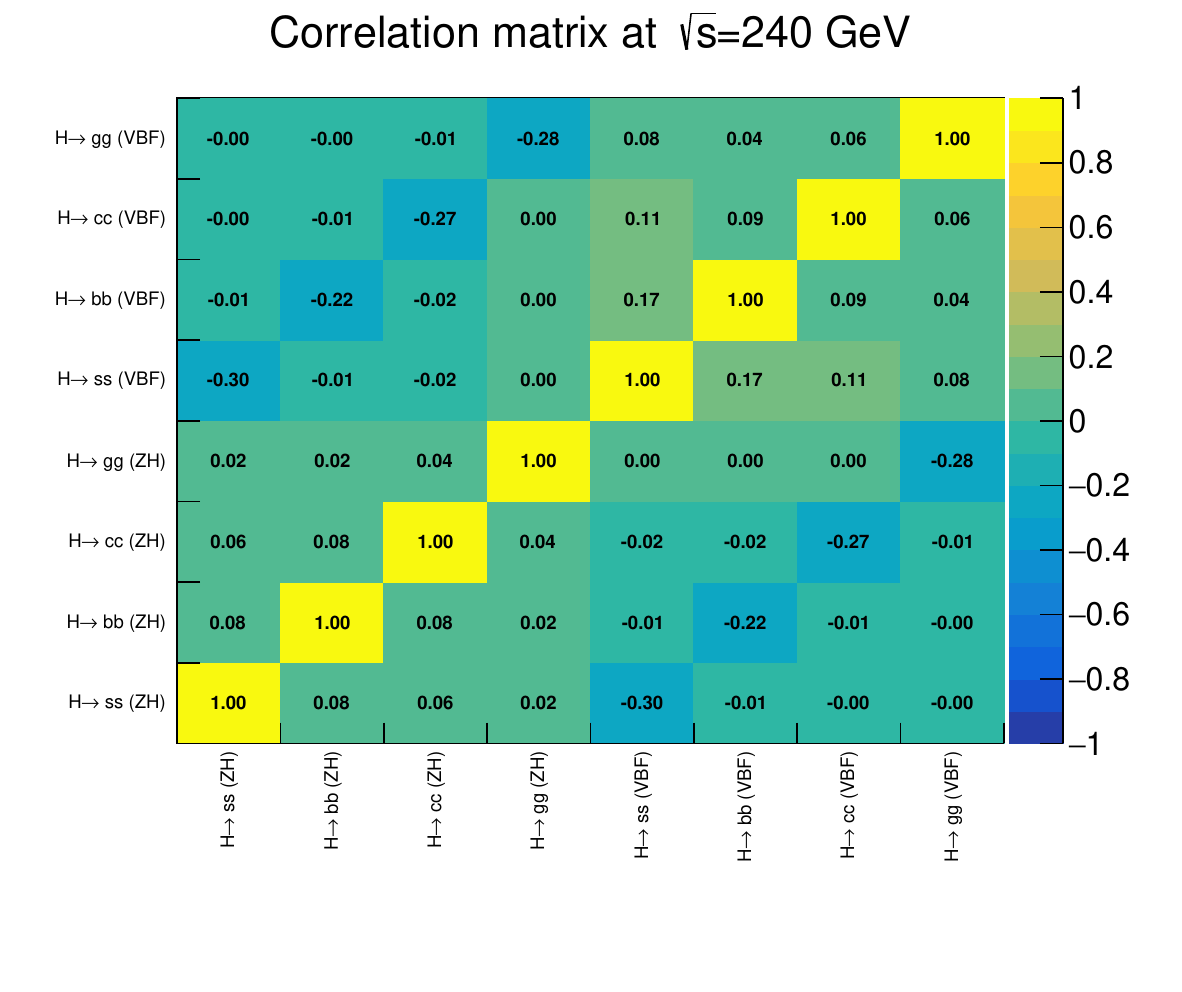}
    \includegraphics[width=0.49\textwidth]{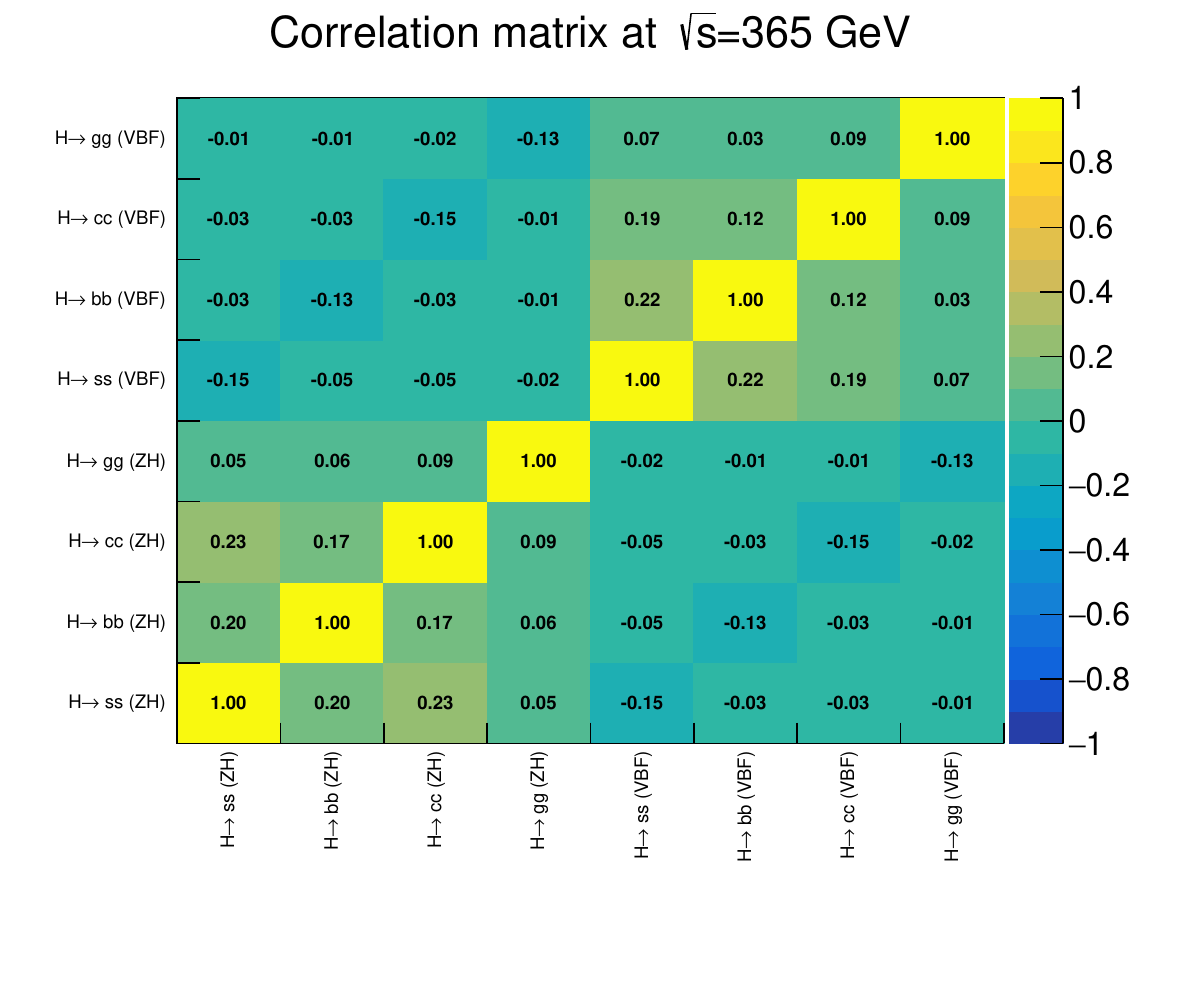}

  \caption{Correlation matrix of the combined fit for $\sqrt{s}=240\,\mathrm{GeV}$ (left) and $\sqrt{s}=365\,\mathrm{GeV}$ (right).
  }
  \label{fig:correlationMatrices}
\end{figure}

The correlation matrices between the fitted signal strengths for the two Higgs boson production and four decay modes under study are shown in Figure~\ref{fig:correlationMatrices} for both $\sqrt{s}=240$ and $365\,\mathrm{GeV}$.

The yields for the $\ell\ell jj$, $\nu\bar{\nu}jj$ and $jjjj$ analysis at $\sqs=365\,\mathrm{GeV}$ are shown in Tables~\ref{tab:llh_365_cutflow}, \ref{tab:pre_selection_365_transformed} and \ref{tab:pre_selection_jjjj_365} respectively.

 \begin{sidewaystable}[!htbp]
\vspace{1cm}
  \centering
  \caption{
    Selection criteria, expected yields and efficiency for signal and background processes for the \texorpdfstring{$\ellell jj$}{lljj} analysis, 
    assuming an integrated luminosity $\mathcal{L}=3.12$~\iab of $e^+e^-$ collisions
    at $\sqs = 365$~GeV.
  }
  \label{tab:llh_365_cutflow}
  \resizebox{\textwidth}{!}{
  \begin{tabular}{lrrrrrrrrrrrrrrrrrrrrrrrr}
\toprule
Selection                   & \multicolumn{2}{c}{$\ell\ell H(b\bar{b})$} & \multicolumn{2}{c}{$\ell\ell H(c\bar{c})$} & \multicolumn{2}{c}{$\ell\ell H(gg)$} & \multicolumn{2}{c}{$\ell\ell H(s\bar{s})$} & \multicolumn{2}{c}{$\ell\ell H(WW)$} & \multicolumn{2}{c}{$\ell\ell H(ZZ)$} & \multicolumn{2}{c}{$\ell\ell H(\tau\tau)$} & \multicolumn{2}{c}{$ZZ$} & \multicolumn{2}{c}{$WW$} & \multicolumn{2}{c}{$t\bar{t}$} & \multicolumn{2}{c}{$Z/\gamma^*(\ell\ell)$} & \multicolumn{2}{c}{$Z/\gamma^*(q\bar{q})$}  \\    
                             & Yield & \effpct &  Yield & \effpct &  Yield  & \effpct &  Yield & \effpct &  Yield & \effpct &  Yield & \effpct & Yield & \effpct &  Yield   & \effpct &  Yield     & \effpct &  Yield    & \effpct &  Yield     & \effpct \\
\midrule
No cuts                      & 21032 &      -  &  1044  &      -  &   2956  &      -  &     8  &     -   &   7771 &      -  &    957 &      -  &  2271 &      -  &  2005536 &       - &  33435480  &      -  &  2496000  &       - &  11909040  &       - &  51224441  &       - \\
$>0$ iso-leptons with $p>40$~GeV & 20710 &     98  &  1028  &     98  &   2911  &     98  &     8  &    98   &   7672 &     99  &    940 &     99  &  2237 &     99  &   371476 &      19 &  12313369  &     37  &   634365  &      25 &   9760686  &      82 &     56782  &       0 \\
one $Z(\ell\ell)$ candidate  & 19255 &     93  &   953  &     93  &   2695  &     93  &     8  &    93   &   7155 &     93  &    876 &     93  &  2103 &     93  &   271476 &      73 &    763372  &      6  &    94082  &      15 &   6235314  &      64 &     41193  &      73 \\
$m_{\ell\ell}$ in 81--101~GeV & 12780 &     66  &   629  &     66  &   1775  &     66  &     5  &    66   &   4757 &     66  &    595 &     68  &  1391 &     67  &   167446 &      62 &     66290  &      9  &    12271  &      13 &   2941287  &      47 &        26  &       0 \\
$m_{\text{recoil}}$ 120--200~GeV & 11468 &     90  &   569  &     90  &   1609  &     91  &     5  &    91   &   4239 &     89  &    518 &     87  &  1241 &     89  &    37196 &      22 &     24020  &     36  &      541  &       4 &    286317  &      10 &         0  &       0 \\
$E(j_2)>15$~GeV             & 11435 &    100  &   568  &    100  &   1608  &    100  &     5  &   100   &   4028 &     95  &    472 &     91  &   941 &     76  &    34910 &      94 &     22461  &     94  &      537  &      99 &    203444  &      71 &         0  &       0 \\
$\leq2$ iso. leptons         & 10522 &     92  &   561  &     99  &   1597  &     99  &     5  &   100   &   2536 &     63  &    383 &     81  &   482 &     51  &    30989 &      89 &     22324  &     99  &      495  &      92 &    203444  &     100 &         0  &       0 \\
$d_{23}>0$, $d_{34}>0$     & 10522 &    100  &   561  &    100  &   1597  &    100  &     5  &   100   &   2532 &    100  &    364 &     95  &   466 &     96  &    25410 &      82 &     11856  &     53  &      495  &     100 &      4980  &       2 &         0  &       0 \\
\midrule
$\ell=e$                    &  5223 &      -  &   278  &      -  &    789  &      -  &     2  &     -   &   1254 &      -  &    180 &      -  &   232 &      -  &    13179 &       - &      6269  &      -  &      253  &       - &      3780  &       - &         0  &       - \\
$\ell=\mu$                  &  5299 &      -  &   283  &      -  &    808  &      -  &     2  &     -   &   1278 &      -  &    185 &      -  &   235 &      -  &    12231 &       - &      5587  &      -  &      243  &       - &      1199  &       - &         0  &       - \\
\bottomrule
  \end{tabular}
  } 
\end{sidewaystable}

\begin{table}[!htbp]
  \centering
  \footnotesize
  \caption{Yields after each step of the selection and final efficiency for the $\nu\bar{\nu}jj$ analysis at $\sqrt{s}=365$~GeV.}
  \begin{tabular}{l|cccc|c}
    \hline
      & Yield & Yield & Yield after & Yield after & Efficiency (\%) \\
    Process & before selection & after lepton veto & $|\cos\theta_{\mathrm{inv}}|<0.85$ & kinematics $\&$ $d_{ij}$ & \\
    \hline
    $\nu_e\bar{\nu}_eH(gg)$       & $7.46\times10^{3}$ & $7.46\times10^{3}$ & $7.43\times10^{3}$ & $5.47\times10^{3}$ & 73.3 \\
    $\nu_e\bar{\nu}_eH(s\bar{s})$ & $2.18\times10^{1}$ & $2.18\times10^{1}$ & $2.18\times10^{1}$ & $1.60\times10^{1}$ & 73.5 \\
    $\nu_e\bar{\nu}_eH(c\bar{c})$ & $2.63\times10^{3}$ & $2.63\times10^{3}$ & $2.54\times10^{3}$ & $1.81\times10^{3}$ & 68.7 \\
    $\nu_e\bar{\nu}_eH(b\bar{b})$ & $5.92\times10^{4}$ & $5.92\times10^{4}$ & $5.30\times10^{4}$ & $3.52\times10^{4}$ & 59.5 \\
    $\nu_e\bar{\nu}_eH(\tau\tau)$ & $4.99\times10^{3}$ & $4.99\times10^{3}$ & $3.52\times10^{3}$ & $6.44\times10^{2}$ & 12.9 \\
    $\nu_e\bar{\nu}_eH(ZZ)$       & $2.28\times10^{3}$ & $2.28\times10^{3}$ & $1.98\times10^{3}$ & $9.66\times10^{2}$ & 42.4 \\
    $\nu_e\bar{\nu}_eH(WW)$       & $2.11\times10^{4}$ & $2.11\times10^{4}$ & $1.48\times10^{4}$ & $8.49\times10^{3}$ & 40.1 \\
    $ZH,\, H(gg)$       & $6.33\times10^{3}$ & $6.33\times10^{3}$ & $6.30\times10^{3}$ & $5.46\times10^{3}$ & 86.3 \\
    $ZH,\, H(s\bar{s})$ & $1.86\times10^{1}$ & $1.86\times10^{1}$ & $1.85\times10^{1}$ & $1.61\times10^{1}$ & 86.6 \\
    $ZH,\, H(c\bar{c})$ & $2.23\times10^{3}$ & $2.23\times10^{3}$ & $2.12\times10^{3}$ & $1.83\times10^{3}$ & 82.0 \\
    $ZH,\, H(b\bar{b})$ & $4.50\times10^{4}$ & $4.50\times10^{4}$ & $3.84\times10^{4}$ & $3.30\times10^{4}$ & 73.3 \\
    $ZH,\, H(\tau\tau)$ & $4.26\times10^{3}$ & $4.26\times10^{3}$ & $2.90\times10^{3}$ & $1.40\times10^{3}$ & 32.8 \\
    $ZH,\, H(ZZ)$       & $1.93\times10^{3}$ & $1.93\times10^{3}$ & $1.65\times10^{3}$ & $1.12\times10^{3}$ & 58.1 \\
    $ZH,\, H(WW)$       & $1.61\times10^{4}$ & $1.61\times10^{4}$ & $1.11\times10^{4}$ & $8.49\times10^{3}$ & 52.7 \\
    $qqH$                            & $1.03\times10^{5}$ & $1.03\times10^{5}$ & $8.20\times10^{4}$ & $9.69\times10^{2}$ & 0.94 \\
    $Z/\gamma^* \to q\bar{q}$        & $2.69\times10^{7}$ & $2.69\times10^{7}$ & $2.68\times10^{7}$ & $2.75\times10^{3}$ & $0.0$ \\
    $WW$                              & $3.26\times10^{7}$ & $3.26\times10^{7}$ & $1.99\times10^{7}$ & $7.29\times10^{5}$ & 2.2 \\
    $ZZ$                              & $1.91\times10^{6}$ & $1.91\times10^{6}$ & $1.44\times10^{6}$ & $2.15\times10^{5}$ & 11 \\
    $t\bar{t}$                        & $2.50\times10^{6}$ & $2.50\times10^{6}$ & $1.36\times10^{6}$ & $1.54\times10^{4}$ & 0.62 \\
    \hline
  \end{tabular}
  \label{tab:pre_selection_365_transformed}
\end{table}

\begin{table}[!htbp]

  \caption{Expected event yields and efficiencies (in \%)  for signal and background processes, after each step of the $jjjj$ event selection, 
    assuming an integrated luminosity $\mathcal{L}=3.12$~\iab of $e^+e^-$ collisions at $\sqrt{s}=365\,\mathrm{GeV}$. Here, $Z(q\bar{q})$ in $ZH$ processes denotes $Z$ boson decays to $u\bar{u}$ and $d\bar{d}$.} \label{tab:pre_selection_jjjj_365} 
    \resizebox{\textwidth}{!}{
        \begin{tabular}{l|ccccc|c} \hline
					& Yield & Yield & Yield after & Yield after & Yield after&  \\ 

        Process & before & after & $m_{\mathrm{vis}},\, \theta_{\mathrm{vis}}$ & $d_{\mathrm{ij}}$ & $\chi^2$ & efficiency (\%) \\ 
					 & selection & lepton veto & requirement &  requirement & requirement &  \\ \hline

        $e^+e^- \rightarrow Z(c\bar{c})H(gg)$ & $2.74\times10^{3}$ & $2.61\times10^{3}$ & $2.07\times10^{3}$ & $2.05\times10^{3}$ & $1.98\times10^{3}$ & 72.3 \\
        $e^+e^- \rightarrow Z(c\bar{c})H(s\bar{s})$ & $6.70\times10^{0}$ & $6.40\times10^{0}$ & $5.09\times10^{0}$ & $5.04\times10^{0}$ & $4.89\times10^{0}$ & 73.0 \\
        $e^+e^- \rightarrow Z(c\bar{c})H(c\bar{c})$ & $9.68\times10^{2}$ & $8.75\times10^{2}$ & $7.17\times10^{2}$ & $7.11\times10^{2}$ & $6.79\times10^{2}$ & 70.1 \\
        $e^+e^- \rightarrow Z(c\bar{c})H(b\bar{b})$ & $1.95\times10^{4}$ & $1.58\times10^{4}$ & $1.34\times10^{4}$ & $1.33\times10^{4}$ & $1.23\times10^{4}$ & 63.1 \\
        $e^+e^- \rightarrow Z(q\bar{q})H(gg)$ & $6.27\times10^{3}$ & $6.22\times10^{3}$ & $4.70\times10^{3}$ & $4.65\times10^{3}$ & $4.57\times10^{3}$ & 72.9 \\
        $e^+e^- \rightarrow Z(q\bar{q})H(s\bar{s})$ & $1.53\times10^{1}$ & $1.53\times10^{1}$ & $1.16\times10^{1}$ & $1.15\times10^{1}$ & $1.13\times10^{1}$ & 73.9 \\
        $e^+e^- \rightarrow Z(q\bar{q})H(c\bar{c})$ & $2.21\times10^{3}$ & $2.09\times10^{3}$ & $1.66\times10^{3}$ & $1.64\times10^{3}$ & $1.59\times10^{3}$ & 72.0 \\
        $e^+e^- \rightarrow Z(q\bar{q})H(b\bar{b})$ & $4.46\times10^{4}$ & $3.79\times10^{4}$ & $3.14\times10^{4}$ & $3.12\times10^{4}$ & $2.93\times10^{4}$ & 65.7 \\
        $e^+e^- \rightarrow Z(b\bar{b})H(gg)$ & $3.49\times10^{3}$ & $3.02\times10^{3}$ & $2.51\times10^{3}$ & $2.49\times10^{3}$ & $2.34\times10^{3}$ & 67.0 \\
        $e^+e^- \rightarrow Z(b\bar{b})H(s\bar{s})$ & $8.53\times10^{0}$ & $7.44\times10^{0}$ & $6.18\times10^{0}$ & $6.14\times10^{0}$ & $5.77\times10^{0}$ & 67.7 \\
        $e^+e^- \rightarrow Z(b\bar{b})H(c\bar{c})$ & $1.23\times10^{3}$ & $1.01\times10^{3}$ & $8.55\times10^{2}$ & $8.50\times10^{2}$ & $7.90\times10^{2}$ & 64.2 \\
        $e^+e^- \rightarrow Z(b\bar{b})H(b\bar{b})$ & $2.48\times10^{4}$ & $1.80\times10^{4}$ & $1.57\times10^{4}$ & $1.56\times10^{4}$ & $1.41\times10^{4}$ & 56.9 \\
        $e^+e^- \rightarrow Z(s\bar{s})H(gg)$ & $3.52\times10^{3}$ & $3.50\times10^{3}$ & $2.67\times10^{3}$ & $2.64\times10^{3}$ & $2.59\times10^{3}$ & 73.6 \\
        $e^+e^- \rightarrow Z(s\bar{s})H(s\bar{s})$ & $8.60\times10^{0}$ & $8.59\times10^{0}$ & $6.58\times10^{0}$ & $6.49\times10^{0}$ & $6.40\times10^{0}$ & 74.4 \\
        $e^+e^- \rightarrow Z(s\bar{s})H(c\bar{c})$ & $1.24\times10^{3}$ & $1.18\times10^{3}$ & $9.37\times10^{2}$ & $9.28\times10^{2}$ & $8.98\times10^{2}$ & 72.4 \\
        $e^+e^- \rightarrow Z(s\bar{s})H(b\bar{b})$ & $2.51\times10^{4}$ & $2.13\times10^{4}$ & $1.78\times10^{4}$ & $1.76\times10^{4}$ & $1.65\times10^{4}$ & 65.7 \\
        $e^+e^- \rightarrow Z(b\bar{b})H(\tau\tau)$ & $2.67\times10^{3}$ & $1.48\times10^{3}$ & $1.32\times10^{3}$ & $1.14\times10^{3}$ & $5.23\times10^{2}$ & 19.6 \\
        $e^+e^- \rightarrow Z(c\bar{c})H(\tau\tau)$ & $2.10\times10^{3}$ & $1.28\times10^{3}$ & $1.17\times10^{3}$ & $9.96\times10^{2}$ & $4.69\times10^{2}$ & 22.4 \\
        $e^+e^- \rightarrow Z(s\bar{s})H(\tau\tau)$ & $2.70\times10^{3}$ & $1.72\times10^{3}$ & $1.58\times10^{3}$ & $1.35\times10^{3}$ & $6.37\times10^{2}$ & 23.6 \\
        $e^+e^- \rightarrow Z(q\bar{q})H(\tau\tau)$ & $4.80\times10^{3}$ & $3.06\times10^{3}$ & $2.81\times10^{3}$ & $2.40\times10^{3}$ & $1.13\times10^{3}$ & 23.5 \\
        $e^+e^- \rightarrow Z(b\bar{b})H(Z\gamma)$ & $6.54\times10^{1}$ & $5.13\times10^{1}$ & $4.08\times10^{1}$ & $3.59\times10^{1}$ & $3.06\times10^{1}$ & 46.8 \\
        $e^+e^- \rightarrow Z(c\bar{c})H(Z\gamma)$ & $2.82\times10^{2}$ & $2.44\times10^{2}$ & $1.90\times10^{2}$ & $1.65\times10^{2}$ & $1.43\times10^{2}$ & 50.7 \\
        $e^+e^- \rightarrow Z(s\bar{s})H(Z\gamma)$ & $6.59\times10^{1}$ & $5.95\times10^{1}$ & $4.56\times10^{1}$ & $3.91\times10^{1}$ & $3.43\times10^{1}$ & 52.1 \\
        $e^+e^- \rightarrow Z(q\bar{q})H(Z\gamma)$ & $1.17\times10^{2}$ & $1.06\times10^{2}$ & $8.05\times10^{1}$ & $6.91\times10^{1}$ & $6.05\times10^{1}$ & 51.7 \\
        $e^+e^- \rightarrow Z(b\bar{b})H(WW)$ & $9.18\times10^{3}$ & $5.31\times10^{3}$ & $4.57\times10^{3}$ & $4.42\times10^{3}$ & $3.36\times10^{3}$ & 36.6 \\
        $e^+e^- \rightarrow Z(c\bar{c})H(WW)$ & $7.21\times10^{3}$ & $4.60\times10^{3}$ & $3.87\times10^{3}$ & $3.72\times10^{3}$ & $2.89\times10^{3}$ & 40.1 \\
        $e^+e^- \rightarrow Z(s\bar{s})H(WW)$ & $9.25\times10^{3}$ & $6.18\times10^{3}$ & $5.09\times10^{3}$ & $4.87\times10^{3}$ & $3.82\times10^{3}$ & 41.3 \\
        $e^+e^- \rightarrow Z(q\bar{q})H(WW)$ & $1.65\times10^{4}$ & $1.10\times10^{4}$ & $8.99\times10^{3}$ & $8.61\times10^{3}$ & $6.73\times10^{3}$ & 40.8 \\
        $e^+e^- \rightarrow Z(b\bar{b})H(ZZ)$ & $1.13\times10^{3}$ & $8.23\times10^{2}$ & $6.70\times10^{2}$ & $6.06\times10^{2}$ & $4.46\times10^{2}$ & 39.5 \\
        $e^+e^- \rightarrow Z(c\bar{c})H(ZZ)$ & $8.85\times10^{2}$ & $7.15\times10^{2}$ & $5.77\times10^{2}$ & $5.18\times10^{2}$ & $3.87\times10^{2}$ & 43.7 \\
        $e^+e^- \rightarrow Z(s\bar{s})H(ZZ)$ & $1.14\times10^{3}$ & $9.62\times10^{2}$ & $7.68\times10^{2}$ & $6.86\times10^{2}$ & $5.16\times10^{2}$ & 45.3 \\
        $e^+e^- \rightarrow Z(q\bar{q})H(ZZ)$ & $2.02\times10^{3}$ & $1.71\times10^{3}$ & $1.36\times10^{3}$ & $1.21\times10^{3}$ & $9.09\times10^{2}$ & 45.0 \\
        $e^+e^- \rightarrow \nu\bar{\nu}H$ & $1.18\times10^{5}$ & $9.98\times10^{4}$ & $2.16\times10^{3}$ & $9.23\times10^{2}$ & 0.0 & $ 0.0$ \\
        $e^+e^- \rightarrow W^+W^-$ & $2.29\times10^{7}$ & $1.46\times10^{7}$ & $9.58\times10^{6}$ & $7.14\times10^{6}$ & $6.39\times10^{6}$ & 27.9 \\
        $e^+e^- \rightarrow ZZ$ & $1.38\times10^{6}$ & $1.06\times10^{6}$ & $5.20\times10^{5}$ & $4.52\times10^{5}$ & $4.13\times10^{5}$ & 29.9 \\
        $e^+e^- \rightarrow Zq\bar{q}$ & $4.91\times10^{7}$ & $4.89\times10^{7}$ & $1.53\times10^{7}$ & $7.81\times10^{6}$ & $7.57\times10^{6}$ & 15.4 \\
        $e^+e^- \rightarrow t\bar{t}$ & $1.86\times10^{6}$ & $9.94\times10^{5}$ & $9.38\times10^{5}$ & $9.18\times10^{5}$ & $7.04\times10^{5}$ & 37.8 \\
        \hline
    \end{tabular}
    }
\end{table}
\acknowledgments

This work was funded in part by the U.S. Department of Energy, Office of Science, Office of High Energy Physics under Contract No. DE-SC0012704.
We would like to thank our colleague Louis Portales for generating several Monte Carlo samples used in these analyses.

\FloatBarrier

\bibliographystyle{JHEP}
\bibliography{main}

\end{document}